\documentclass[twocolumn]{aastex631}
\usepackage[whole]{bxcjkjatype} 
\usepackage{comment}
\usepackage[normalem]{ulem}
\usepackage{amsmath}

\newcommand{\cm}{{\mathrm{\,cm}}}

\newcommand{\second}{{\mathrm{\,s}}}

\newcommand{\km}{{\mathrm{\,km}}}
\newcommand{\kmNSP}{{\mathrm{km}}}
\newcommand{\kms}{{\km\second^{-1}}}
\newcommand{\kmsNSP}{{\kmNSP\second^{-1}}}

\newcommand{\au}{{\mathrm{\,au}}}

\newcommand{\pc}{{\mathrm{\,pc}}}

\newcommand{\yr}{{\mathrm{\,yr}}}

\newcommand{\numden}{{\cm^{-3}}}

\newcommand{\msunNSP}{{M_\sun}}
\newcommand{\msun}{{\,\msunNSP}}

\newcommand{\rsun}{{R_\sun}}
\newcommand{\solarmass}{\msun}
\newcommand{\solarmassyr}{{\solarmass\yr^{-1}}}

\newcommand{\K}{{\mathrm{\,K}}}

\defcitealias{Shu_1991}{SRLL}
\defcitealias{Shang_2020}{I}
\defcitealias{Shang_2023}{II}
\defcitealias{Shang_PIII}{III}

\begin{document}

\title{A Unified Model for Bipolar Outflows from Young Stars: Kinematic and Mixing Structures in HH 30}

\author[0000-0002-0517-1197]{Tsung-Han Ai （艾宗瀚）}
\affiliation{Institute of Astronomy and Astrophysics, Academia Sinica, Taipei 10617, Taiwan}
\affiliation{Department of Physics, National Central University, 300 Zhongda Road, Zhongli, Taoyuan 32001, Taiwan}

\author[0000-0002-1624-6545]{Chun-Fan Liu（劉君帆）}
\affiliation{Institute of Astronomy and Astrophysics, Academia Sinica, Taipei 10617, Taiwan}

\author[0000-0001-8385-9838]{Hsien Shang （尚賢）}
\affiliation{Institute of Astronomy and Astrophysics, Academia Sinica, Taipei 10617, Taiwan}

\author[0000-0002-6773-459X]{Doug Johnstone}
\affiliation{NRC Herzberg Astronomy and Astrophysics, 5071 West Saanich Rd, Victoria, BC, V9E 2E7, Canada}
\affiliation{Department of Physics and Astronomy, University of Victoria, Victoria, BC, V8P 5C2, Canada}

\author[0000-0001-5557-5387]{Ruben Krasnopolsky}
\affiliation{Institute of Astronomy and Astrophysics, Academia Sinica, Taipei 10617, Taiwan}

\footnote{T.-H. Ai and C.-F. Liu contribute equally to this work.}
\correspondingauthor{Hsien Shang, shang@asiaa.sinica.edu.tw}

\begin{abstract}
The young stellar source HH 30 is a textbook example of an ionic optical jet originating from a disk in an edge-on system shown by the HST. It has a remnant envelope in $^{12}$CO observed by ALMA. The optical jet is characterized by its narrow appearance, large line width at the base, and high temperature inferred from line diagnostics. Three featured structures can be identified, most evident in the transverse position--velocity diagrams: an extremely--high-velocity (EHV) wide-angle wind component with large spectral widths in the optical, a very--low-velocity (VLV) ambient surrounding medium seen in $^{12}$CO, and a low-velocity (LV) region traced by $^{12}$CO nested both in velocity and location between the primary wind and ambient environment. A layered cavity with multiple shells forms nested morphological and kinematic structures around the optical jet. The atomic gas originating from the innermost region of the disk attains a sufficient temperature and ionization to emit brightly in forbidden lines as an optical jet. The wide-angle portion expands, forming a low-density cavity. The filamentary $^{12}$CO encompassing the wind cavity is mixed and advected inward through the action of the magnetic interplay of the wide-angle wind with the molecular ambient medium. The magnetic interplay results in the layered shells penetrating deeply into the vast cavity of tenuous atomic wind material. The HH 30 system is an ideal manifestation of the unified wind model of \citet{Shang_2020,Shang_2023}, with clearly distinguishable atomic and molecular species mixed through the atomic lightly ionized magnetized wind and the surrounding cold molecular ambient material.

\end{abstract}


\section{Introduction} \label{sec:introduction}

Outflows play significant roles in star formation. Yet, how they form remains an open question. The material ejected from the adjacent vicinity of the protostar sweeps up the ambient gas and produces a low-density cavity enclosed by compressed high-density shells \citep[see, e.g.,][]{Shu_1991,Liang_2020,Rabenanahary_2022}. Highly collimated jets are the extremely--high-velocity \citep[EHV, $v\gtrsim50\kms$; see, e.g.,][]{Bally_2016} component of an outflow system, which are accompanied by wide and narrow cavities. Previously, the observational resolution has limited the understanding of the detailed structures within these outflows. Furthermore, the structures between the jet and the cavity wall are difficult to detect without having high spatial resolution and high sensitivity. 

Recent advancement in resolution and sensitivity for telescopes such as the Atacama Large Millimeter/submillimeter Array (ALMA) and James Webb Space Telescope (JWST) enables the detection of the complex interior structures of these outflow lobes. Nested structures of outflows from embedded Class 0/I sources have been revealed at sub-arcsecond resolutions by ALMA \citep[e.g., HH 211, HH 212, DG Tau A, DG Tau B, and HH 30;][]{Jhan_2021,Lee_2022,deValon_2022,Louvet_2018} and JWST \citep[e.g., L1527,\footnote{\url{https://webbtelescope.org/contents/news-releases/2022/news-2022-055}} TMC1A, IRAS 15398$-$3359;][]{Harsono_2023,Yang_2022}, allowing for detailed kinematic studies and confrontation with high-resolution theory and numerical simulation results. Yet, even before the era of ALMA and JWST, the launch of the Hubble Space Telescope (HST) revolutionized the field of star formation by providing, for the first time, high--angular-resolution ($\sim0\farcs1$) imaging of bright optical protostellar jets in terms of Herbig--Haro (HH) objects \citep[e.g.,][and references therein]{Hartigan_2011,Erkal_2021}.

HH 30 is associated with a nearby ($d\sim140\pc$) low-mass young stellar object (YSO) in the Taurus star-forming region. Revealed by the HST, 
HH 30 shows a textbook protostellar system with a pair of jet and counterjet bright in optical forbidden lines and two reflection nebulae divided by a dark lane indicating an almost exactly edge-on disk \citep{Burrows_1996,Ray_1996}. The HH 30 YSO is classified as Class I by its spectral energy distribution \citep[SED,][]{Wood_2002,Furlan_2008,Madlener_2012}. Modeling the reflection nebulae suggests that a flattened disk-like structure, with only an additional tenuous envelope, explains the observed features \citep{Wood_1998}. The northeastern portion of the disk surface is facing toward the observer with an angle $\sim88^\circ$ from the line of sight \citep{Burrows_1996, Wood_1998}, harboring the approaching jet. Proper motion studies of the northeastern jet spanning 20 years yield an average tangential velocity of $\sim100\kms$ \citep{Estalella_2012}, corresponding to the average jet velocity for the edge-on HH 30.

The optical jet of HH 30 has been a target of interest for HST since its debut with the Wide Field and Planetary Camera 2 (WFPC2). \citet{Burrows_1996} examined the $R$-band and $I$-band images and derived an increasing trend of jet image width with a decreasing trend of intensity, suggestive of expanding flow characteristics. \citet{Bacciotti_1999} derived the physical conditions along the jet using HST and ground-based narrow-band images based on an inversion technique from \citet{BE99} that utilizes several optical forbidden line ratios.
A gradually decreasing trend of electron density along the flow following the electron recombination curve was found starting from $n_e\gtrsim10^6\numden$ at within $0\farcs2$, with a relatively stable ionization fraction $x_e\approx0.1$, and a temperature $T_e$ decreasing from $\sim2\times10^4\K$ down to $\sim6\times10^3\K$ beyond $\sim2\arcsec$. \citet{Hartigan_2007} extended the derivation to simultaneously match various forbidden emission line ratios between 4000 and 8000 \AA\ extracted from the Space Telescope Imaging Spectrograph (STIS) line-integrated ``slitless'' spectroscopy, revealing downstream as well as lateral variations of physical conditions in the jet. \citet{Coffey_2008} inferred average values of physical condition $n_e\approx1.5\times10^4\numden$, $T_e\approx6000\K$, and $x_e\approx0.1$ at $0\farcs6$ from the star. The HH 30 optical jet is thus known to be hot and partially ionized.

The molecular content of this Class I source reveals the parent cloud material inherited into the system. Using the James Clerk Maxwell Telescope (JCMT), \citet{Moriarty-Schieven_2006} examined the molecular environment around HH 30 and its nearby forming stars.
Subsequent observations by Plateau de Bure Interferometer (PdBI) in molecular line emission, including $^{12}$CO, C$^{18}$O, HCO$^{+}$, and continuum at 2.7 and 3.3 mm, uncovered the kinematic information of the gaseous disk and the V-shaped outflow lobe \citep{Pety_2006}. C$^{18}$O indicated a counterclockwise Keplerian rotating disk. The monopolar northeastern $^{12}$CO outflow lobe extending up to $5\arcsec$ above the disk with an opening angle of $\sim35^{\circ}$ is suggested to trace the conical shell morphology. ALMA's sub-arcsecond resolution and high-sensitivity observations revealed more kinematic details of its conical $^{12}$CO outflow \citep{Louvet_2018}. Ellipse fitting applied on the transverse position--velocity diagrams (PVDs) from close to the disk plane to $250 \au$ above the disk shows the gas flowing along the cavity surface with constant velocity $v = 9.3 \kms$. \citet{Louvet_2018} compared the V-shaped morphology present in the $^{12}$CO outflow to a photoevaporated wind \citep{Alexander_2014} or a magnetocentrifugal disk wind with a launching radii ranging possibly from $0.5$ to $2.5\au$ \citep{Ferreira_2006}.
Both mechanisms, however, would not be able to produce the hot and ionized jet of $100\kms$ observed in the optical forbidden line emission.

The X-wind model \citep{Shu_1994_1,Shu_1995_5}, magnetocentrifugally launched from the innermost edge of the circumstellar disk, has been proposed to explain the highly collimated appearance and large line widths observed by optical forbidden emission lines \citep{Shang_1998}.
Emission line diagnostics of X-winds were obtained from the thermal properties and ionization conditions self-consistently calculated in \citet{Shang_2002,Shang_2010}, which regards X-ray irradiation resulting from stellar magnetic activity as the most crucial ionization source. The produced temperature, ionization, and forbidden emission line intensities and ratios align with optical and infrared jets from low-mass YSOs, e.g., [\ion{O}{1}], [\ion{S}{2}], [\ion{Ne}{2}], and [\ion{Ne}{3}]. 

The unified wind model with a jet-bearing wide-angle X-wind, which treats the outflow system around a YSO as a highly magnetized bubble, has been proposed to explain the observed kinematic structures, including the EHV jets, winds, multi-cavities, and episodic shells \citep{Shang_2006,Shang_2020,Shang_2023}. Nested kinematic and morphological features in Class 0/I molecular outflows, revealed by ALMA observations, such as HH 212, HH 211, CARMA-7, and DG Tau B, have been discussed and accounted for under the framework of the X-wind driven wind bubbles by \citet{Shang_2020,Shang_2023}. This paper aims to provide an integrated picture of the observational atomic and molecular line data of HH\,30 obtained at optical and radio wavelengths using the unified model for bipolar outflows made by \citet{Shang_2020,Shang_2023}. 
The detailed structure of a young molecular outflow, which shows nested and layered filamentary structures within the cavities, may be attributed to the magnetic interplay between an atomic and partially ionized wide-angle X-wind and the molecular ambient material.

The paper is organized as follows. Section \ref{section:data sets} provides an overview of the archival data obtained from HST (Section \ref{section:HST data}) and ALMA (Section \ref{section:ALMA data}), including details about the analyzed region where moment 0 maps and transverse and parallel PVDs were constructed. Section \ref{section:Results of Data} presents the observed features in the HST and ALMA datasets. The kinematic properties of the HH 30 jet/outflow system are discussed in Section \ref{section:Discussion}. Section \ref{section:Atomic wind} presents the evidence of the atomic wind signature manifesting an X-wind jet, the basic principles of the X-wind model, and the associated parameters for the HH 30 optical jet.
Section \ref{section:CO as Mix molecular gas} discusses the ambient nature of the $^{12}$CO gas. Section \ref{section:HH30_illustration} demonstrates how HH 30 is an illustrative example of the magnetic interplay of the atomic X-wind and the molecular ambient material. Section \ref{section:Louvet} discusses the disk wind scenarios explored, e.g., by \citet{Louvet_2018}, and Section \ref{section:diff_dw} is dedicated to the difficulties of the 
disk wind scenarios to a system such as HH 30. Section \ref{section:summary} summarizes our findings and interpretations.

\section{Analyzed Data Sets}
\label{section:data sets}

Our analysis investigates previously observed optical and radio data sets from the HST and ALMA archives. The HST imaging and spectroscopic data were individually taken in 2003 and 2008, whereas this ALMA spectroimaging data cube was observed in 2015.

\subsection{HST Data}
\label{section:HST data}

\subsubsection{HST/STIS Multiple-Slit Spectra along the Jet Axis}
\label{section:HST data Multiple-Slit}

HH 30 was observed by the Hubble Space Telescope Imaging and Spectrograph (HST/STIS) under the General Observing (GO) Program 9164 (PI: Wiseman), and the data were obtained from the Mikulski Archive for Space Telescopes (MAST). The spectra taken in January 2003 are analyzed here. The dataset consists of spectra taken from multiple slit positions placed along the jet axis, at position angle $31\fdg65$ and stepping across the jet from $-0\farcs32$ to $0\farcs24$ along the position angle of $121\fdg65$ with a step of $0\farcs08$. For each slit position, two dithers $0\farcs25$ apart were performed along the slit direction, and the two dithered spectra were shifted to match the physical positions and averaged. Due to guide star failure during the visit, only eight spectra (4 locations) centered from 0 to $0\farcs24$ have sufficient signal-to-noise ratios for further analysis. The \texttt{G750M} grating and the \texttt{52X0.1} slit were used, resulting in a wavelength coverage of 6295 to 6867 \AA\ and a velocity resolution of ${\sim}45\kms$ sampled at ${\sim} 25\kms\,{\rm pixel^{-1}}$. Atomic lines from [\ion{O}{1}] $\lambda\lambda6300,6363$, [\ion{N}{2}] $\lambda\lambda6548,6583$, H$\alpha$, and [\ion{S}{2}] $\lambda\lambda6716,6731$ were covered in the spectra.

The setup of the multi-slit observations produces a spectral datacube covering $\sim0\farcs34$ across the southeastern side of the jet axis. It allows for analyses in terms of PVDs parallel and transverse to the jet axis. The spectral datacube was first median-combined across the jet axis to form effective ``coadded'' spectral images of the aforementioned emission lines as \textit{parallel} PVDs. Along the jet axis, at every $\sim0\farcs25$ from the nominal source of the HH 30 jet, a range of $\sim0\farcs25$ is averaged to form a series of \textit{transverse} PVDs.
   
\subsubsection{Archival HST/WFPC2 Broad-Band Imaging}
\label{section:HST data broad-band}

An HST/Wide Field and Planetary Camera 2 (WFPC2) image of HH 30 (GO 11867, PI: Trauger) was retrieved via enhanced data products from the Hubble Legacy Archive (HLA). The observation was conducted in 2008 October 26 and November 10. The HLA-produced data combined with cosmic-ray rejection through the standard pipeline using \texttt{MultiDrizzle}. HH 30 was placed on the PC chip of the four chips used by WFPC2, with a Nyquist-limited pixel size of $0\farcs05$. The broad-band imaging was observed using the \texttt{F675W} filter, including the $R$-band forbidden lines of [\ion{O}{1}], [\ion{N}{2}], and [\ion{S}{2}], as well as the permitted H$\alpha$ emission line. The same filter setup has been used for the data taken in 1994 and 1995, with a coarser pixel sampling of $0\farcs1$ on the WF2 chip, and presented by \citet{Burrows_1996}. Compared with the HST/WFPC2 narrow-band imaging reported by \citet{Ray_1996}, the jet features shown in the broad-band image mainly reflect those observed in the forbidden emission lines of [\ion{O}{1}] and [\ion{S}{2}].

\subsection{Archival ALMA Data} 
\label{section:ALMA data}

We obtained radio datacubes of HH 30 from the ALMA archive (project ID: 2013.1.01175.S, PI: Dougados). The $^{12}$CO emission line at 230.538 GHz was observed by Band 6 on July 19, 2015, at an angular resolution of $0\farcs228$ and with a field of view $25\farcs820$ centered at $(\alpha,\delta)=$ (04:31:37.469, +18:12:24.220) (J2000).
ALMA pipeline-reduced $^{12}$CO and $^{13}$CO datacubes were retrieved, and the beam sizes are $0\farcs26 \times 0\farcs22$ and $0\farcs27 \times 0\farcs23$, respectively. 
For the jet/outflow axis, we adopt a position angle of $31\fdg3$, in line with the value of $31\fdg2\pm0\fdg1$ adopted by \citet{Louvet_2018}. The disk axis is defined to be perpendicular to the jet axis. 

After obtaining the given angles and positions, we utilized our own Python code, which includes libraries such as Scipy, OpenCV, and SpectralCube, to customize the range of each PVD to aid in the analysis. Parallel and transverse PVDs are obtained from the datacube by integrating the $^{12}$CO emission across the jet and disk axes to investigate the spatial evolution of kinematic structures within the material emanating from the source star.
The parallel PVD includes the whole lobe width of $10\arcsec$ across the jet axis. The transverse PVDs are formed consecutively at different heights, every $0\farcs25$ along the northeastern outflow lobe from close to the disk midplane to $\sim3\farcs75$ above the disk. An effective ``slit'' width of $0\farcs25$ is used for averaging.
The systemic velocity of HH 30 is defined \citep{Louvet_2018} by the peak of the intensity located at $6.90\kms$ in the $^{13}$CO line profile.

\section{Results of Data analysis}
\label{section:Results of Data}

\subsection{Observed Features of Optical Forbidden Emission}
\label{section:Results_Optical}

\subsubsection{Proper Motions and Knot Identifications}

The archival HST and ALMA data, spanning epochs between 2003 and 2015, are used in combination for the panoramic examination of both the atomic and molecular content of the HH 30 jet/outflow system. Although exact knot tracing is beyond the scope of this work, knot identifications and known proper motions are worth noting. The optical broad-band image of HH 30 includes the forbidden emission lines that trace the collimated structure along the jet axis. The brightest knot in the image is identified as nearly stationary over 20 years, designated as 01N in the space-borne HST image \citep{Burrows_1996} and A0 in ground-based images \citep{Anglada_2007,Estalella_2012}. The optically unseen HH 30 stellar position was measured to be shifted by $0\farcs51$ from the A0 knot \citep{Burrows_1996}, and we adopt that separation to set the origin of the driving source for comparison of optical and radio data in this work.

Table \ref{tab:knot_proper motion} summarizes the proper motions adopted from \citet{Estalella_2012} and expected positions at various observational epochs of the brightest knots within $\sim10\arcsec$ from the source, identified as A0, A1, A2, and A3 following the nomenclature of \citet{Anglada_2007} and \citet{Estalella_2012}. 
The proper motions of these knots (the A-group knots A1 through A3) are found to be roughly $0\farcs15\yr^{-1}$ to $0\farcs18\yr^{-1}$ \citep[corresponding to tangential velocity of $\sim100\kms$ based on a $140\pc$ distance;][]{Estalella_2012}.
Since HH 30 is nearly edge-on, the tangential velocities of the knots, $\sim100\kms$, may be regarded as the flow velocity.
These proper motions result in $\sim0\farcs7$ displacement between the 2003 and 2008 epochs and $\sim1\farcs1$ displacement between 2008 and 2015. 

Figure \ref{fig:HST_PPV} shows the HST/WFPC2 $R$-band image and HST/STIS spectra of optical forbidden emission lines, with the expected knot positions summarized in Table \ref{tab:knot_proper motion} indicated by blue horizontal bars. The knots can be identified up to A2, whereas the A3 knot becomes too diffuse to be detected by HST. Compared with the ground-based [\ion{S}{2}] imaging in \citet{Estalella_2012}, an additional knot is resolved from A0 in the HST image and spectra, at $\sim1\farcs1$ and $\sim1\farcs4$ from the driving source in the 2003 spectra and 2008 image, respectively. This might be a newly formed knot after the 1999 observation of \citet{Anglada_2007}, which we tentatively give a ``Aa1'' knot designation. Its proper motion appears to be roughly half that of A1, and its confirmation requires follow-up high-resolution imaging.

\begin{deluxetable*}{ccccccc}
\tabletypesize{\scriptsize}
\tablecolumns{5}
\tablewidth{0pt} 
\tablecaption{Estimation of the A-group knot positions based on proper motions \label{tab:knot_proper motion}}
\tablehead{
    \colhead{       } & \multicolumn{3}{c}{\citet{Estalella_2012}} & \colhead{HST/STIS Spectra} & \colhead{HST/WFPC2 Imaging} & \colhead{ALMA Band 6}\\
    \colhead{       } & \multicolumn{3}{c}{2010} & \colhead{2003} & \colhead{2008} & \colhead{2015}\\
    \colhead{knot} & \colhead{$v_t$ ($\kms$)} & \colhead{$\mu_{y}$ ($\arcsec\yr^{-1}$)} & \colhead{$y$ (\arcsec)} & \multicolumn{3}{c}{expected $y$ (\arcsec)}
}
\startdata
{A0} & \nodata & \nodata & $ 0.51$ & $\sim 0.51$ & $\sim 0.51$ & $\sim 0.51$\\
{Aa1}& \nodata & \nodata & \nodata & $\sim 1.10$ & $\sim 1.40$ & \nodata\\
{A1} & 101.5 & 0.153 & $ 3.20$ & $2.13$ & $2.89$ & $3.97$\\
{A2} & 99.50 & 0.150 & $ 4.63$ & $3.58$ & $4.33$ & $5.38$\\
{A3} & 118.7 & 0.179 & $ 7.36$ & $6.11$ & $7.00$ & $8.26$\\
\enddata
\tablecomments{The knot positions in 2010 were extracted from \citet{Estalella_2012}, with the A0 knot serving as the reference point. To match the coordinates used in our study, the heights reported here were adjusted, with the A0 knot positioned at approximately $0\farcs51$ above the disk \citep{Burrows_1996,Anglada_2007,Estalella_2012}. Knot positions listed in the archival data are estimated values from individual proper motions, except for the newly resolved knot Aa1 in HST images and spectra after 2002.}
\end{deluxetable*}

\subsubsection{Overview of Broad-Band Imaging}

\begin{figure*}[ht!]
\plotone{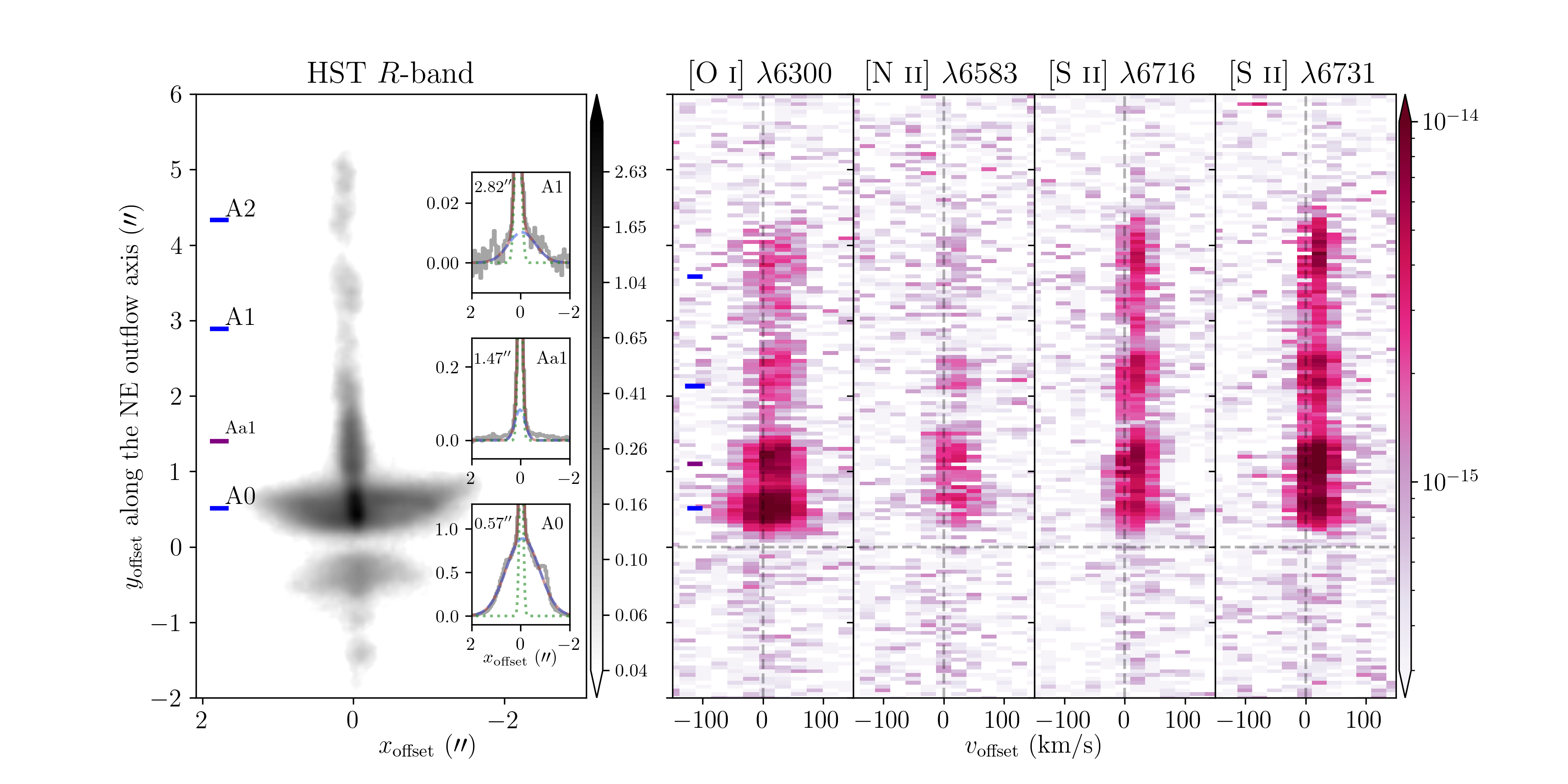}
\caption{HST/WFPC2 $R$-band image (left) and HST/STIS parallel PVDs of HH 30 along the jet axis at a position angle of $31\fdg65$, shown in [\ion{O}{1}] $\lambda6300$, [\ion{N}{2}] $\lambda6583$, and [\ion{S}{2}] $\lambda\lambda6716,6731$ lines (right). The velocities shown are relative to the systemic velocities of $v_{\rm helio} = 20.1\kms$ \citep{Appenzeller_2005}. The blue horizontal bars are expected knot positions of A0, A1, and A2 at observation epochs of 2008 (the WFPC2 image) and 2003 (the STIS spectra) summarized in Table \ref{tab:knot_proper motion}. The additional purple bars are the newly resolved knot, Aa1, between A0 and A1, unresolved in the 2010 ground-based imaging \citep{Estalella_2012}. The insets in the left panel show the spatial profiles at selected positions roughly corresponding to knots A0 ($\sim0\farcs57$), Aa1 ($\sim1\farcs47$), and A1 ($\sim2\farcs82$) and their corresponding two-Gaussian fitting profiles. The observed spatial profiles are shown in gray. The fitted narrow jet component is shown in a green dotted line, the fitted broad nebular component is in a blue dashed line, and the combined fitted curve is in a red line. The heights of the insets have been restricted to 1/10 of the peak intensities to show the broader component.
\label{fig:HST_PPV}}
\end{figure*}

The left panel of Figure \ref{fig:HST_PPV} shows the optical $R$-band image of HH 30 obtained by the HST. The broad-band image contains both the $R$-band continuum as well as emission lines. As noted by \citet{Burrows_1996}, the continuum is dominant in the reflection nebula and the jet and counterjet are bright in the emission lines. The brightest knot, A0, is enveloped by one side of the reflection nebula.

Examination of the image across the jet axis suggests that the emission consists of two spatial components of different intensities. The components can be separated by fitting double Gaussians to the spatial profiles. The brighter component has a typical spatial full-width at half-maximum (FWHM) of $\sim0\farcs2$ that can be well fitted by a Gaussian, corresponding to the optical jet. 
The HST marginally resolves this apparent jet width with roughly twice its nominal spatial resolution of $0\farcs1$.
The other component is typically fainter by a factor of $\sim5$ and  in the image becomes unobservable beyond $\sim\pm3\arcsec$ from the nominal stellar position. This component
has a spatial FWHM of $\sim1\arcsec$ to $2\arcsec$ with a skewed shape deviating from a Gaussian profile, corresponding to the reflection nebula. The insets in the left panel of Figure \ref{fig:HST_PPV} show the spatial profiles at selected knot positions, roughly at A0, Aa1, and A1, along with the two-Gaussian fitting results.

\subsubsection{Properties of Optical Forbidden Emission along the Jet Axis}

The right panels of Figure \ref{fig:HST_PPV} show the coadded parallel PVDs of the forbidden emission lines of [\ion{O}{1}] $\lambda6300$, [\ion{N}{2}] $\lambda6548$, and [\ion{S}{2}] $\lambda\lambda6716,6731$. These lines can be traced down to $\sim0\farcs1$ from the star, at which the optically thick disk starts to block the direct and scattered stellar light. The line centroids appear to be redshifted by no more than a pixel ($\lesssim25\kms$) for the systemic velocity of $v_\mathrm{helio}\approx20.1\kms$ \citep{Appenzeller_2005}. The bright jet being redshifted is consistent with previous ground-based spectroscopy with higher dispersion \citep[e.g.,][]{Anglada_2007,Appenzeller_2005}. A redshifted jet on the approaching side of the modeled disk has been proposed as evidence for a wobbling disk system. This wobbling could be due to the driving source of HH 30 being part of a binary system \citep{Anglada_2007,Estalella_2012}.

The optical forbidden lines mainly trace the knots detected in the optical broad-band image. The brightest A0 ``knot'', considered stationary at the base of the jet, exhibits the largest line widths. At A0, [\ion{O}{1}] exhibits approximately $70\kms$ FWHM ($\gtrsim60\kms$ if deconvolved with the $45$-$\kmsNSP$ instrumental width at a $25$-$\kms$ sampling)
and $\lesssim200\kms$ full-width at zero-intensity (FWZI), while [\ion{S}{2}] has a slightly smaller FWHM of $\sim55\kms$ and FWZI of $\sim150\kms$. The adjacent resolved Aa1 knot, with similar peak intensity, has narrower line widths by $\sim10\%$ in both [\ion{O}{1}] and [\ion{S}{2}]. The further knots A1 and A2 become fainter in [\ion{N}{2}] and [\ion{O}{1}] while maintaining similar intensities in [\ion{S}{2}]. For these knots, the [\ion{O}{1}] line has a FWHM $\sim60\kms$ while [\ion{S}{2}] has a line width $\sim40\kms$ that is only marginally resolved by the medium dispersion grism of STIS\@.

These forbidden emission lines are evidence of a high-density and highly ionized jet from HH 30 that can be traced to within $0\farcs1$ from the star. The [\ion{O}{1}]$\lambda6300$ line is sensitive to electron density at and beyond the scale of $n_{\rm cr}\approx10^6\numden$. The bright [\ion{S}{2}] doublet also traces an ionic region with electron density  $n_e\gtrsim10^4\numden$, with [\ion{N}{2}]$\lambda6583$ indicative of highly ionized gas. The lines are brightest at the first two knots A0 and Aa1, and become fainter downstream, which is most evident in [\ion{O}{1}] and [\ion{N}{2}]. The relative line intensities along the flow suggest that the main ionization and high-density region occurs towards the region close to the star, whereas the outer downstream regions recombine along the flow without major replenishment of ionization and with no density enhancement.

The optical forbidden line ratios reflect the physical conditions of the jet mainly in terms of electron density, temperature, and ionization fraction. 
The first two bright knots A0 and Aa1 have [\ion{S}{2}]$\lambda\lambda6716/6731$ ratios of $\lesssim0.6$, which corresponds to $n_e>10^4\numden$, and the ratio increases to $\gtrsim1$ along the flow as $n_e$ drops down to below $\sim10^3\numden$. The [\ion{N}{2}] $\lambda6583$/[\ion{O}{1}] $\lambda6300$ ratio has a low value of $\lesssim0.1$ at $\sim0\farcs5$ but reaches as high as $\gtrsim1$ along the flow which indicates a highly ionized flow of $x_e>0.1$. The [\ion{S}{2}] $\lambda6731$/[\ion{O}{1}] $\lambda6300$ ratio has reached $\sim0.02$ at A0 and increases up to 0.1 and 1 downstream, which requires at least $T_e$ ranging from $\sim6000\K$ to $\gtrsim10000\K$.

\subsubsection{Properties of Optical Forbidden Emission across the Jet Axis}
\begin{figure*}
\plotone{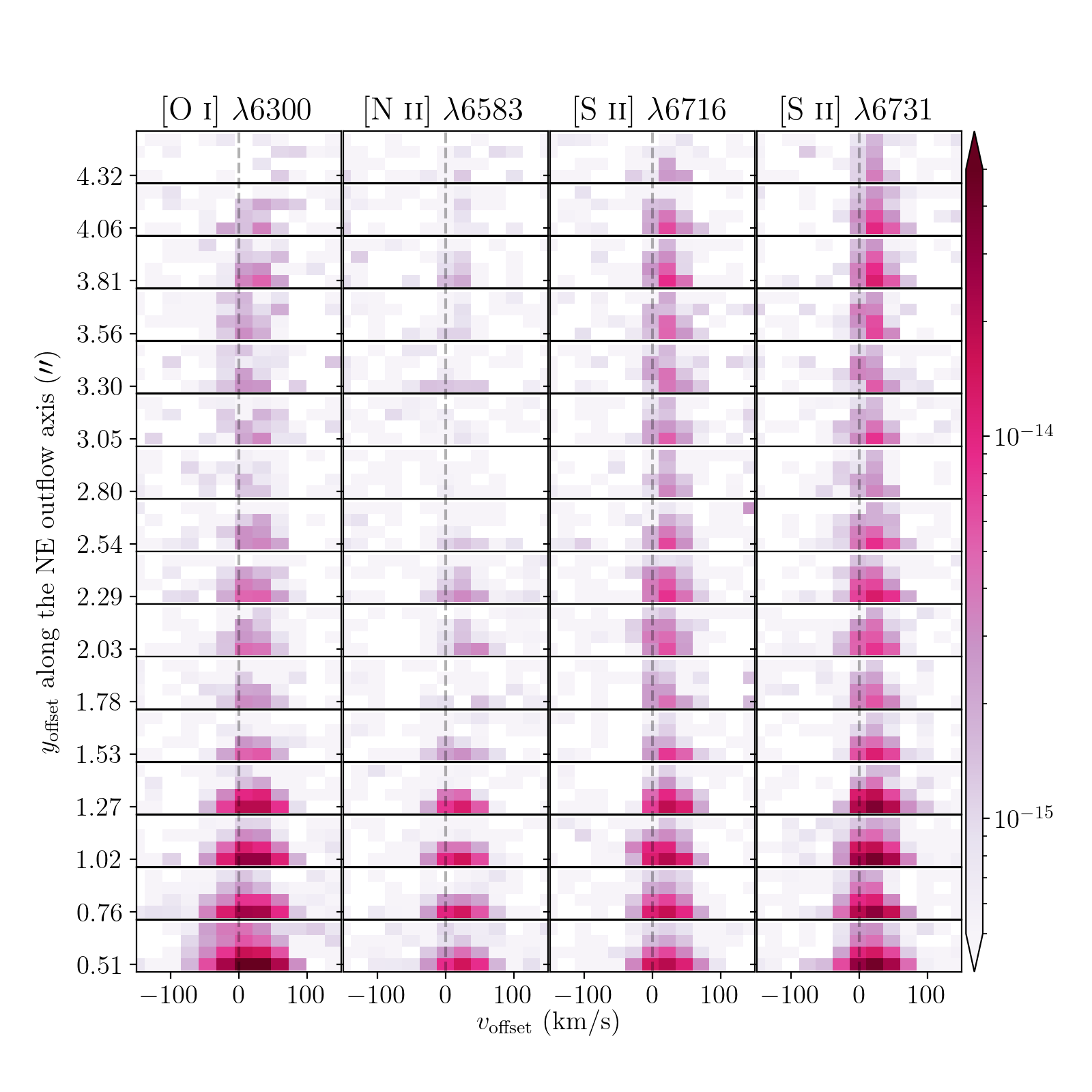}
\caption{Transverse PVDs of HH 30 in the optical forbidden emission lines observed by the HST/STIS GO-9164 program. Each panel shows the portion of a transverse PVD extracted at a specific offset position along the jet axis (labeled on the overall vertical axis), averaged along the jet within a range of $\sim0\farcs25$. For each panel, the radial velocity is shown in the horizontal axis, relative to the systemic $v_{\rm helio}=20.1\kms$ \citep{Appenzeller_2005}, and the transverse distance across the jet axis is shown in vertical axis from $-0\farcs05$ to $0\farcs30$ on the position angle of $121\fdg65$.
    \label{fig:HST_TPV_GO9164}}
\end{figure*}

Figure \ref{fig:HST_TPV_GO9164} shows the reconstructed transverse PVDs from the multiple-slit HST/STIS observation. The 4 forbidden emission lines of [\ion{O}{1}], [\ion{N}{2}], and [\ion{S}{2}], as in Figure \ref{fig:HST_PPV}, are presented. For each emission line, the transverse PVDs were constructed at positions every $0\farcs25$ by averaging within a range of $\sim0\farcs25$ along the jet at various lateral slit positions across the jet. Each position's corresponding transverse PVD is shown with line-of-sight velocity as the horizontal axis and distance across the jet as the vertical axis. Due to failure in positional acquisitions, only half of the jet width was covered.
This causes the emission to be half of an oval shape.

Along the jet flow, the variation of the emission line properties resembles the trends observed in the parallel construction of the PVDs. Both the line widths and line intensities decrease along the jet axis. The variations from $0\farcs51$ to $1\farcs27$ appear relatively continuous, covering the brightest knot, A0, and the resolved Aa1 knot. The more distant A1 ($\sim2\farcs29$) and A2 ($\sim3\farcs56$) knots are much fainter due to lower excitation conditions, most evident in the [\ion{N}{2}] line. 

The transverse PVDs for each position show lateral line intensities and width variations.
For example, at $\sim0\farcs51$ (A0), the optical forbidden lines can be seen to show an FWZI of $\sim200\kms$ on the jet axis and $\sim100\kms$ at $\sim0\farcs25$ away from the jet axis, along with a lateral decrease in line intensities. Overall, the line intensities diminish to the background values at $\sim0\farcs3$  from the axis, consistent with the image FWHM of $\sim0\farcs2$ as derived from the HST broad-band imaging. The lateral changes of line intensities and widths may be mainly an effect of a lateral density decrease such that the most diverging flow streamlines become less detectable when traced cylindrically away from the jet axis. Moreover, the extent of lateral variations also differs among emission lines. The [\ion{N}{2}] line, for example, shows a more compact lateral extent than the [\ion{O}{1}] and [\ion{S}{2}] lines. The overall behavior of line ratios shows greater variation along the flow than across the flow.

\subsection{Observed Features of \texorpdfstring{$^{12}$CO}{12CO} Line Emission}

\subsubsection{Overview of Morphology on Moment 0 and Channel maps}
\label{section:Morphological overview}
\begin{figure*}[ht!]
\plotone{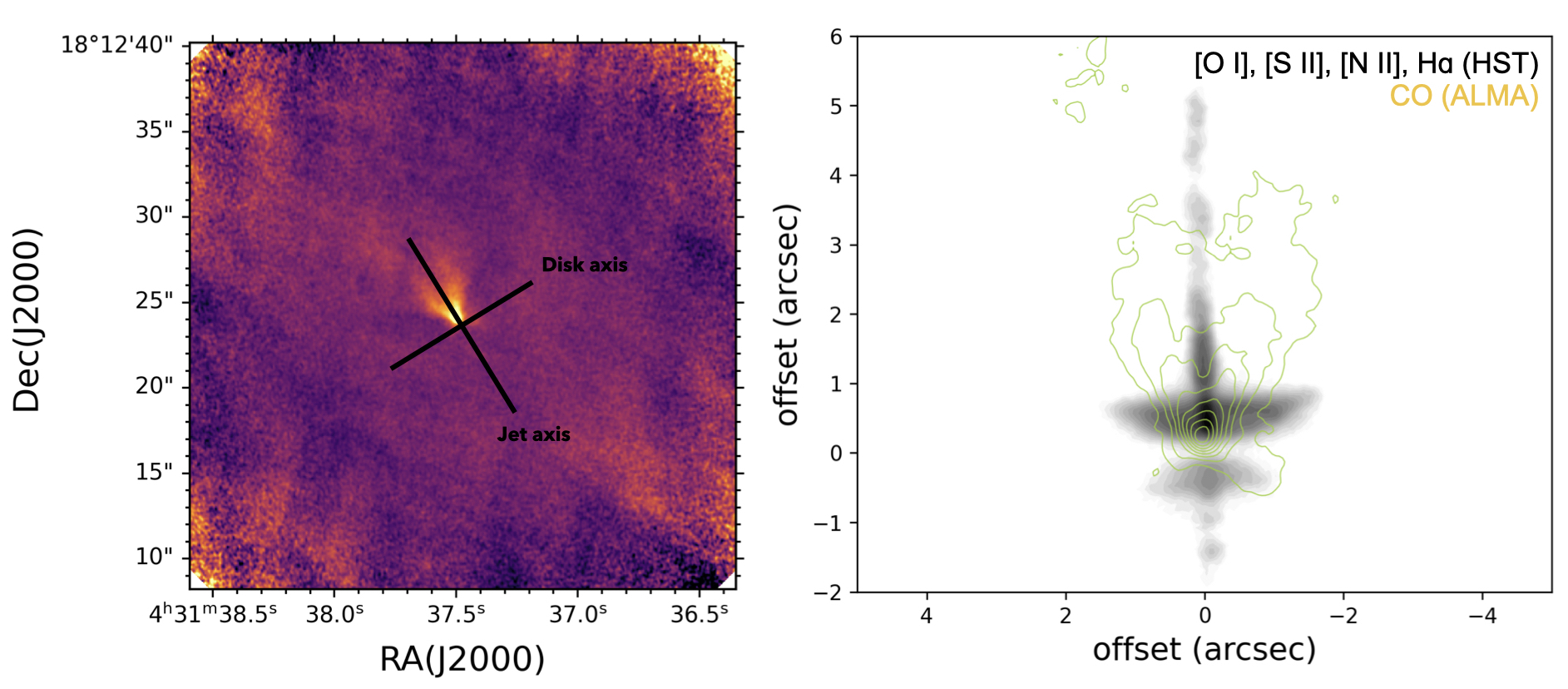}
\caption{\textit{Left:} ALMA $^{12}$CO integrated intensity map over $v_\mathrm{LSR} = -5 \kms$ to $v_\mathrm{LSR} = 20 \kms$ range, with north up; \textit{Right:} optical HST/WFPC2 $R$-band image, including forbidden [\ion{O}{1}], [\ion{S}{2}], [\ion{N}{2}] and permitted H$\alpha$ lines, with ALMA $^{12}$CO integrated intensity map overlaid in green contour, rotated such that the position angle of $31\fdg3$ is pointing up.
\label{fig:M0map}}
\end{figure*}

Multiple components within the outflow, each exhibiting diverse kinematic properties, can be traced by emission lines excited by various physical conditions. $^{12}$CO is utilized to trace the low-velocity (LV) components as presented in the left panel of Figure \ref{fig:M0map}. Moment 0 map of HH 30 integrated from $v_\mathrm{LSR} = -5 \kms$ to $v_\mathrm{LSR} = 20 \kms$ shows a clear monopolar outflow lobe pointing towards the northeast.
On top of the ALMA $^{12}$CO map, the right panel of Figure \ref{fig:M0map} overlaps the optical $R$-band image obtained by the HST\@.
The $^{12}$CO-integrated intensity pattern exhibits a distinct V shape, characterized by a decrease in intensity as one moves away from the source along the outflow axis. The optical jet is located between the arms of the V shape, suggesting the components observed in the optical are well embedded within the $^{12}$CO structure.

\begin{figure*}[ht!]
\plotone{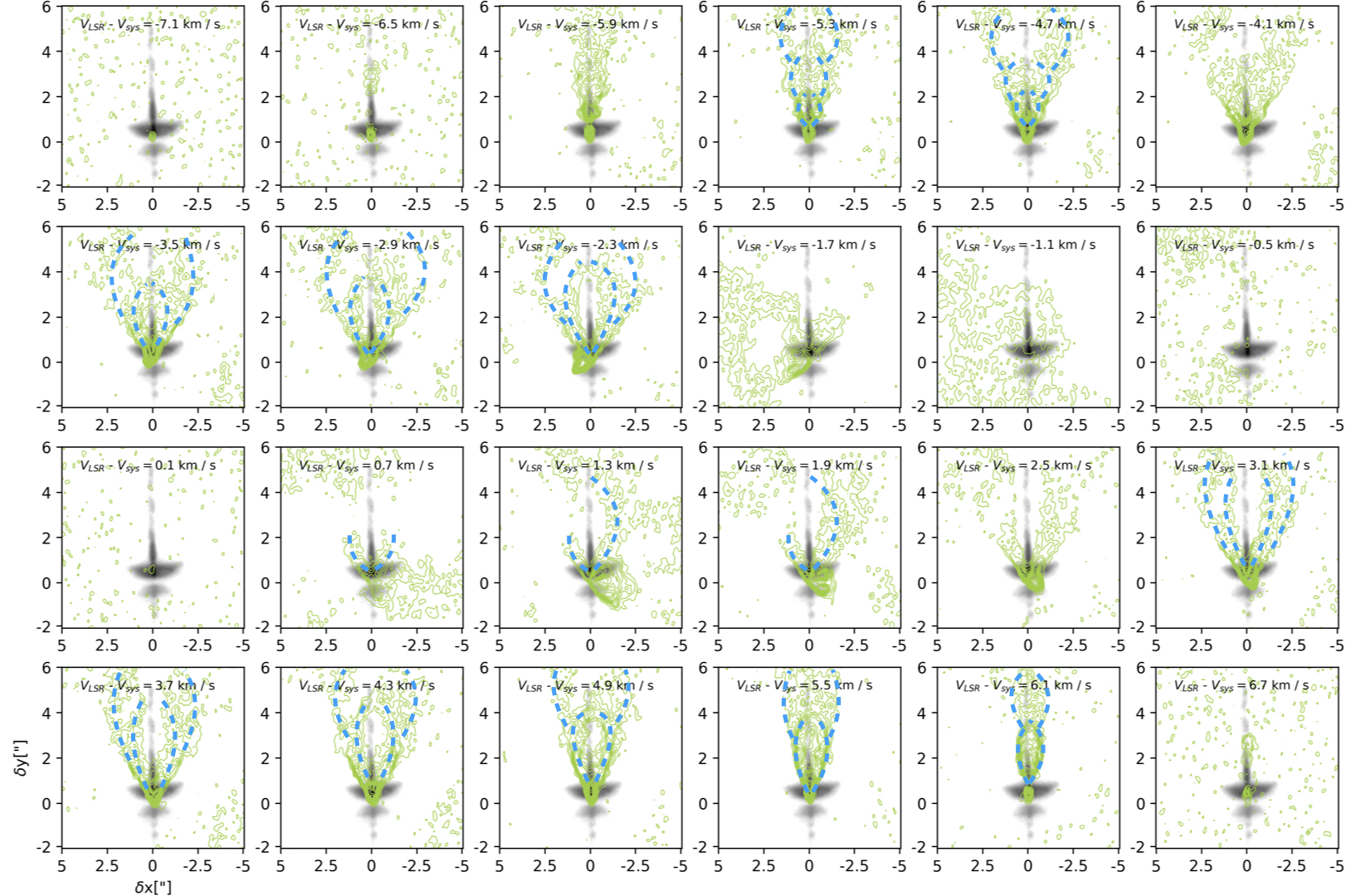}
\caption{ALMA $^{12}$CO channel maps from $v_{\rm LSR}-v_{\rm sys} = -7.1$ to $6.7\kms$ shown as green contours, rotated such that the P.A. of $\sim31\fdg3$ points up. Optical HST/WFPC2 image is overlaid as gray-scale colormap on every channel for comparison. Depending on velocity channels, the droplet-shaped (or V-shaped) $^{12}$CO finger-like arcs are delineated in thick gray dashed lines. The clumps in the most blueshifted/redshifted channels are circled by thin blue dashed lines.
\label{fig:COChMap_HSTImg}}
\end{figure*}

The larger spatial scale of $^{12}$CO compared to the optical forbidden line emission may also be identified through the $^{12}$CO channel maps with more details revealed. Figure \ref{fig:COChMap_HSTImg} presents the $R$-band jet image (as in Figure \ref{fig:M0map}) overlaid on each of the $^{12}$CO channel maps at a channel width of $0.6\kms$. Both optical and $^{12}$CO present bright and dense emission at about $0\farcs5$ above the disk mid-plane. Droplet-shaped finger-like arcs (shown in thick gray dashed lines) encompassing the optical jet are presented in the redder and bluer channels at $v_{\rm LSR}-v_{\rm sys} = -5.3$ to $-2.3\kms$ and $4.3$ to $6.1\kms$. The tips of these finger-like arcs are not clear in the channel close to the systemic velocity, which makes the V-shaped morphology within $v_{\rm LSR}-v_{\rm sys} = -1.7$ to $3.7\kms$.
The two most blueshifted/redshifted channels with $v_{\rm LSR}-v_{\rm sys} = -6.5\kms$ and $6.7\kms$ show clumpy structures concentrating toward the jet axis $\sim 3\arcsec$ above the disk (circled by the thin blue dashed lines).
A trend of decreasing transverse extent of the $^{12}$CO LV component with respect to the increasing line-of-sight velocities can be seen from the channel maps.

\subsubsection{Parallel Position--Velocity Diagrams}
\label{section:Parallel PV}

\begin{figure}[ht!]
\epsscale{1.3}
\plotone{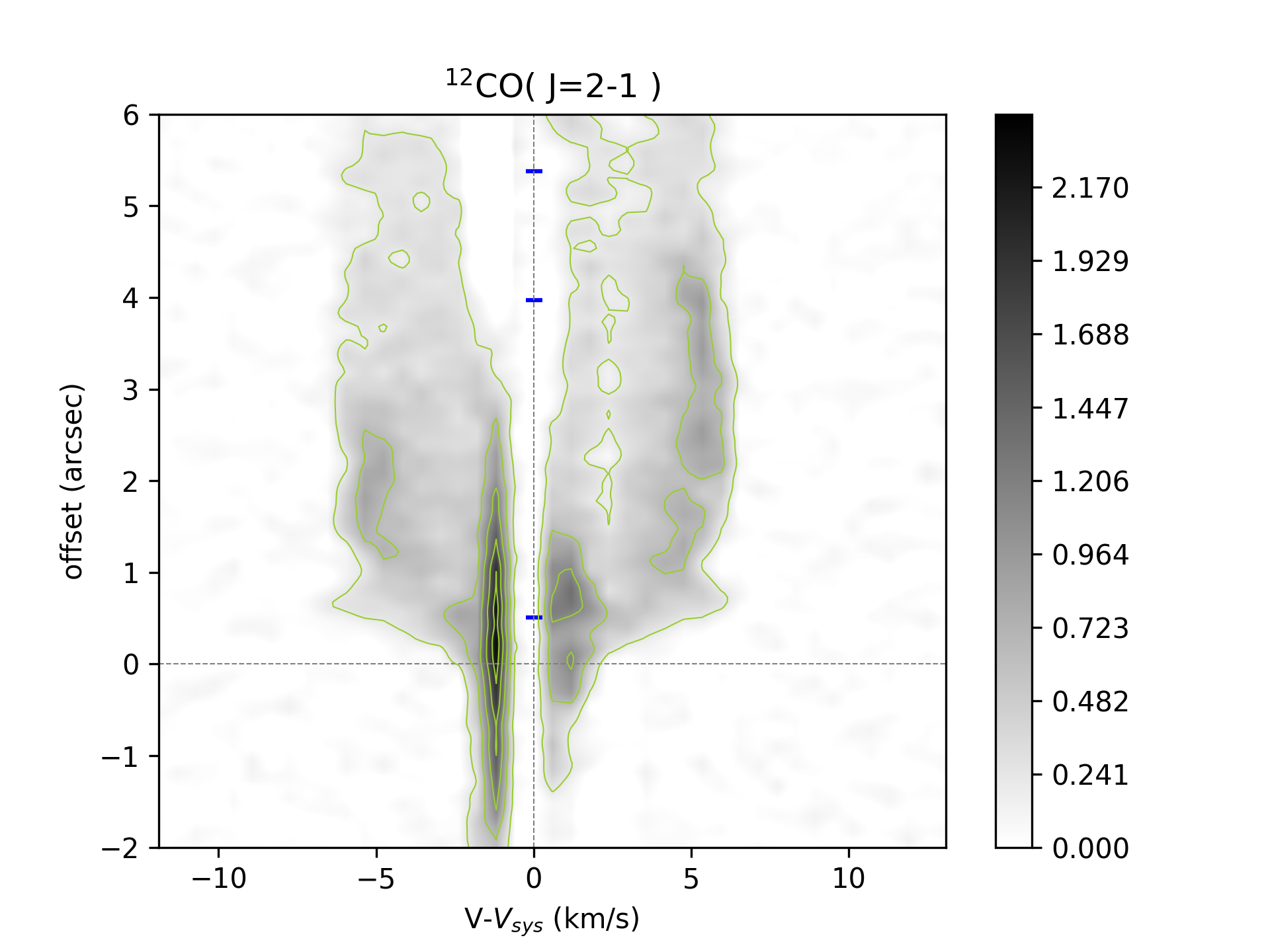}
\caption{ALMA parallel PVD of $^{12}$CO along the position angle of $31\fdg3$. The velocities shown are relative to the systemic velocities of HH 30 of $v_{\rm LSR} = 6.9\kms$ \citep{Louvet_2018}. The expected positions of the optical knots A0, A1, and A2 at the epoch of the 2015 ALMA observation, as shown in Table \ref{tab:knot_proper motion}, are shown in blue horizontal bars.
\label{fig:ALMA_PPV}}
\end{figure}

Parallel PV maps of $^{12}$CO, obtained from integrating across the jet axis, along the jet axis within $-2\arcsec$ to $6\arcsec$ are exhibited in Figure \ref{fig:ALMA_PPV}. 
A pair of prominent peaks on either side of the redshifted/blueshifted part in the line profile at the height of $\sim0\farcs5$ above the disk is observed, consistent with the position of the A0 knot. Very--low-velocity (VLV) components ranging from systemic velocity to $\pm2\kms$ extend from $-2\arcsec$ to $3\arcsec$ along the jet axis. The VLV redshifted components start to fade out over a height of $3\arcsec$ above the disk. Interestingly, there is a potential separation between blueshifted LV and VLV components.

We note clear kinematic differences by comparing the radio and optical emission line features along the jet. 
The optical forbidden emission lines exhibit broader line widths, approximately $\sim\pm100\kms$, compared to the outer lower velocity part of the $^{12}$CO emission at $\sim\pm 5 \kms$ with a width of $\sim2\kms$. Additionally, the velocity dispersion of the optical forbidden emission decreases as the height along the jet axis increases from $\sim0\farcs5$ to $3\arcsec$, which differs from the stable trend observed in the $^{12}$CO emission.

\subsubsection{Transverse Position--Velocity Diagrams}
\label{section:Transverse PV}

\begin{figure*}[ht!]
\plotone{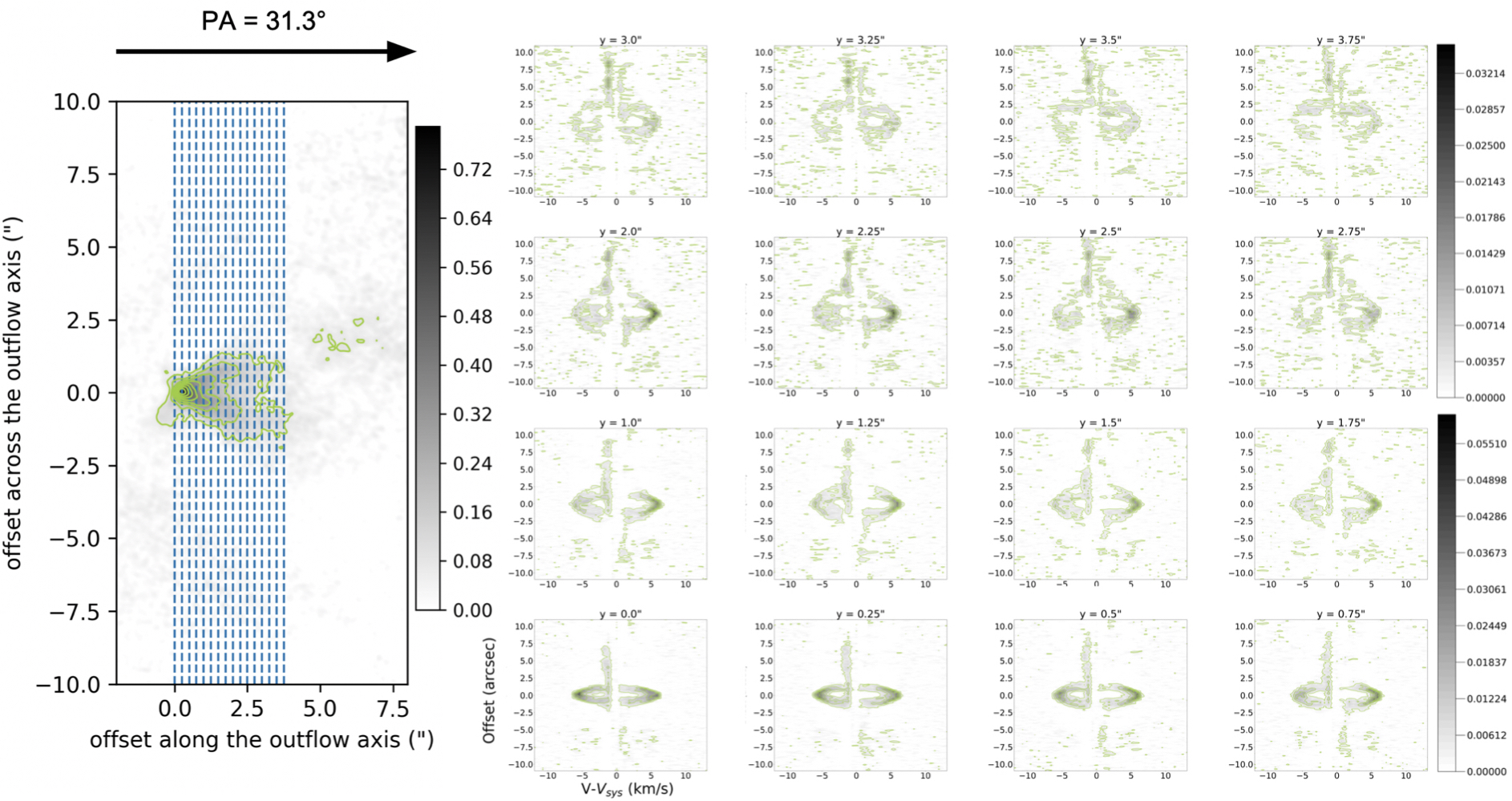}
\caption{Transverse PVDs of HH 30 along disk axis at different heights. Left: $^{12}$CO $(J=2-1)$ moment 0 map shown in greyscale and green contours. Blue dashed lines mark the cut positions of each transverse PVD. The axes are rotated to fit the position axis of the transverse PVDs on the right. Right: Transverse PVDs integrating the flux within a $0\farcs25$ width at heights from the disk plane to $3\farcs75$ above the disk. 
\label{fig:TPV_m0_and_clips}}
\end{figure*}

Transverse PVDs of $^{12}$CO are shown in Figure \ref{fig:TPV_m0_and_clips}. These are made to investigate the spatial and, thus, temporal evolution of kinematic structures within the outflow emanating from the source. Cuts are shown at different heights above the disk plane up to $3\farcs75$ with a $0\farcs25$ width for each transverse PV map.

\begin{figure}[ht!]
\plotone{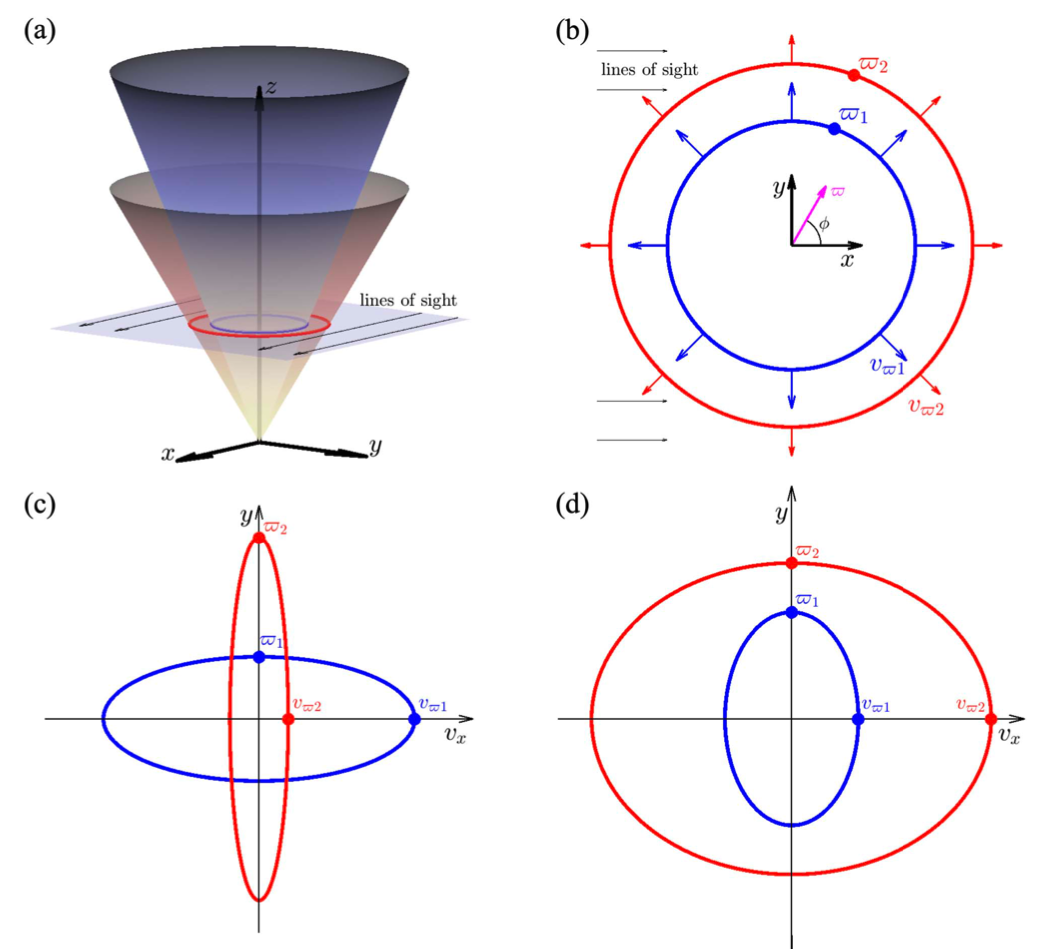}
\caption{Kinematic structure of a simplified two-shell outflow model: (a) schematics of two simplified cone-shape shells with constant radial velocities viewed edge-on; 
(b) an edge-on slice cutting through the two-shell outflow coplanar to the line of sight at a constant height, with projected velocity vectors, indicated; cases of transverse PVDs with the line-on-sight velocity in the horizontal axis, showing (c) the lower-velocity shell being the outer one, and (d) vice versa. Adopted from \citet{Shang_2023}.
\label{fig:Twoshell_model}}
\end{figure}

\begin{figure}[ht!]
\epsscale{1.2}
\plotone{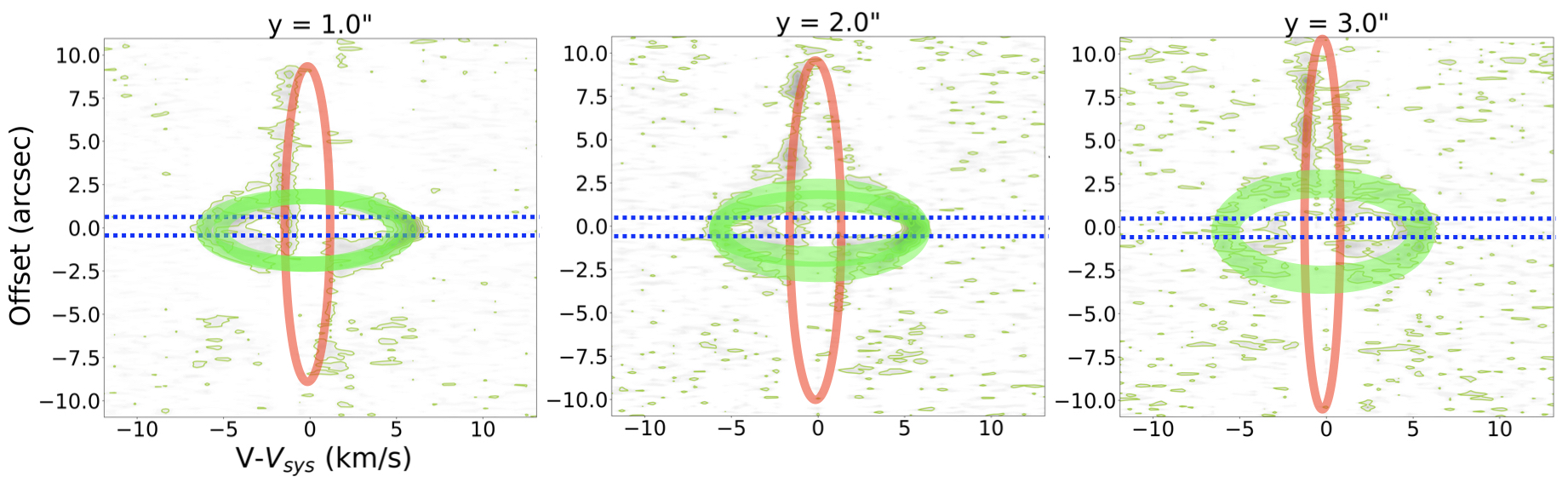}
\caption{Transverse PVDs, at the height of $1\farcs0$, $2\farcs0$ and $3\farcs0$ above the disk, extracted from Figure \ref{fig:TPV_m0_and_clips}, revealing the presence of two components observed in CO: VLV (red) and LV (green) ellipses. Additionally, there is a component observed in optical data (in dashed blue) whose velocity scale is ten times larger than that of the $^{12}$CO observations, and it is presented as a mimic pattern in the $^{12}$CO velocity scale. These components are represented by the ellipses marked on the transverse PVDs.
\label{fig:TPV_Sketch of ellipses}}
\end{figure}

Ellipse-like structures are observed on transverse PVDs. Figure \ref{fig:Twoshell_model} exhibits a simplified two-shell model proposed by \citet{Shang_2023} to illustrate the kinematic meaning of the ellipses observed on transverse PVDs of edge-on sources. Based on tomography analysis, \citet{Louvet_2018} also suggested that the elliptical shapes are due to the projection effect of circularly symmetric structures.

Two primary ellipse-like structures as illustrated in Figure \ref{fig:TPV_Sketch of ellipses}, marked by red (VLV) and green (LV) patterns, are distinguishable based on the lateral position axis distribution ($y$-axis) and have high similarity to the case illustrated in Figure \ref{fig:Twoshell_model}(c). In that case, the outer component possesses a lower spatial velocity than the inner component.

The VLV component, divided into two segments around $v \approx \pm 1 \kms$, forms an elliptical ring shape, delineated by the red ovals in Figure \ref{fig:TPV_Sketch of ellipses}.   
While ascending from the disk plane to higher altitudes, this component maintains nearly constant velocity and experiences a slight expansion from $\sim \pm 7 \arcsec$ to $\sim \pm 10\arcsec$ across the jet axis. 
The lateral extent of the VLV component indicates its presence at the outskirts of the outflow lobe and encompasses all the structures.

The LV thick structure shows a nearly constant or slightly decreasing velocity up to $\sim\pm5\kms$ from the disk plane up to a height of $3\farcs75$. It also reveals a consistent expansion at the positional axis, which can be observed over a range of scales, from $\lesssim1\arcsec$ to approximately $5\arcsec$, as shown by the green ovals in Figure \ref{fig:TPV_Sketch of ellipses}.
The LV ellipse actually corresponds with diverse structures at different heights. As shown in Figure \ref{fig:COChMap_HSTImg}, these structures are the bright knots close to the base of the outflow, a droplet-shaped structure at mid-altitudes, and an additional vaguer (droplet-shaped) envelope at higher altitudes that is not recognizable in all channels. 
The morphology revealed by the moment 0 map and PVDs indicates that the LV ellipse traces finger-like arcs and exhibits complex kinematic structures. 
Despite being composed by multiple arcuate structures, they are still axisymmetric on large scales, resulting in an elliptical projection on the PV space.

The inner CO-dark void of the LV ellipse arises at around $0\farcs5$ above the disk.
It initially has a spatial width of approximately $0\farcs25$ across the jet axis. As the height rises to $2\farcs0$ above the disk, the void gradually expands to a width of $2\arcsec$ across the jet axis. From that point, it maintains the same width until it reaches a distance of $3\farcs75$ from the disk plane. 
The characteristics of the LV structure below $0\farcs5$ from the mid-plane, before where the inner void arises, are dominated by the hyperbolic shapes probably manifesting disk features. This is consistent with the fact that the optical emission, as shown in Figures \ref{fig:HST_PPV} and \ref{fig:HST_TPV_GO9164}, is obscured by the disk and does not emerge until beyond $0\farcs5$ from the source.

The structure observed by the optical forbidden lines should also be considered to understand the outflow lobe's complete structures fully. As presented in Figure \ref{fig:HST_TPV_GO9164}, 
the optical emission illustrates a line width of around $200 \kms$ and can be treated as an EHV ellipse laying on the position--velocity space, confined within a narrow region only $<0\farcs35$ across the jet axis. This emission spatially fills in while kinematically extrudes beyond the apparent CO-dark void. 

Notably, the spatial distribution of the optical forbidden lines agrees with the inner CO-dark void observed in the $^{12}$CO transverse PVDs. This implies that the material located within the inner CO-void region, surrounded by the LV ellipse, is primarily atomic. As a result, it is not detectable in CO, but is bright in optical forbidden lines such as [\ion{O}{1}], [\ion{S}{2}], and [\ion{N}{2}]. It corresponds to the axial, collimated jet portion that is dense and hot enough for exciting the forbidden line emission, as discussed in \citet{Shang_2002}. 

By examining the ellipses present in both the $^{12}$CO and optical transverse PVDs, the nested features of the VLV, LV, and EHV ellipses can be shown in Figure \ref{fig:TPV_Sketch of ellipses}, corresponds to  three structures with respectively $\pm1\kms$, $\pm5\kms$ (with width of $\sim 2\kms$), and $\pm100\kms$ from $v_\mathrm{sys}$. The kinematic interpretations and the physical meaning of these features will be discussed in the next Section.

\section{Discussion}
\label{section:Discussion}

This section discusses the significant astrophysical implications of the HST and ALMA data analyzed and interpreted in this work.

\subsection{The Optical Ionized Jet and the Atomic Wind}
\label{section:Atomic wind}

The HST optical imaging and spectroscopic data reveal the physical and kinematic properties of the HH 30 jet. Bright in optical forbidden emission lines, e.g., [\ion{O}{1}], [\ion{S}{2}], and [\ion{N}{2}], the jet is only marginally resolved by the $\sim0\farcs1$ spatial resolution of HST. The constructed transverse PVDs also show a lateral decrease in line intensities.
The obtained line ratios are within the ranges given by previous observations \citep[e.g.,][]{Bacciotti_1999,Hartigan_2007,Coffey_2008}. These observations align with the concept that the jet primarily consists of atomic matter with an ionization fraction of approximately $\sim10\%$. The inner jets may have densities as high as $10^6\numden$ for the [\ion{O}{1}] forbidden lines to be bright and temperatures up to $10^4\K$, \citep[e.g.,][]{BE99,Shang_2002,Liu_2012}. The emission lines possess large widths of $\sim200\kms$ close to the jet base, almost twice the jet velocity of $\sim100\kms$ derived from proper motion measurements, characteristic of a wide-angle wind.

In the X-wind model, the X-wind is launched from the inner disk edge truncated by the star--disk interaction \citep{Shu_1994_1,Shu_1995_4}. The disk truncates at radius $R_\mathrm{x}$ as

\begin{equation}
    R_\mathrm{x} = \Phi_{\rm dx}^{-4/7} \left(\frac{\mu_\ast^4}{GM_\ast\dot{M}_{\rm D}^2}\right)^{1/7},
\end{equation}
where $\Phi_{\rm dx}$ is a factor of order unity, $M_\ast$ is the stellar mass, $\mu_\ast$ is the stellar magnetic dipole moment, and $\dot{M}_\mathrm{D}$ is the disk accretion rate  \citep{Shu_1994_1,Shu_1995_4}. At $R_\mathrm{x}$, the Keplerian velocity has a value 
\begin{equation}
v_\mathrm{x} = R_\mathrm{x} \Omega_\mathrm{x} = R_\mathrm{x} \left(\frac{GM_\ast}{R_\mathrm{x}^3}\right)^{1/2}.
\end{equation}

For the free portion of the X-wind in axisymmetric steady state, which is intrinsic to a cold magnetocentrifugal wind, all field lines (i.e., streamlines) in the jet and the wide-angle wind share the same starting point at $(z,\varpi)=(0, R_\mathrm{x}$) from where they fan out to cover both the jet and the wide-angle wind. Each field line has its specific angular momentum $J_\mathrm{w}$, conserved along each wind streamline, with an average of $\bar{J}_\mathrm{w}$.
The terminal poloidal velocity is $\bar{v}_\mathrm{w} = v_\mathrm{x} \sqrt{2\bar{J}_\mathrm{w}-3}$, where $\Omega_\mathrm{x}R_\mathrm{x}=v_\mathrm{x}$ at the launch location of the inner edge of the disk, because all the wind streamlines originate from the same innermost $\mathrm{X}$-region.
This is unique to the X-wind, where all streamlines 
originate at the same $v_\mathrm{x}$. The X-wind can be easily determined once the direct source parameters 
$\bar{v}_\mathrm{w}$, $M_\ast$, $\mu_\ast$ and $R_\mathrm{x}$ are given.

The stellar mass of HH 30 has been dynamically determined as $0.45\msun$ \citep{Pety_2006}, and the stellar radius is adopted to be $3\rsun$ \citep{Shu_1994_1,Shu_1996,Shu_1997}. The stellar field strength is not directly inferred. The $R_\mathrm{x}$ can be dynamically obtained with the Keplerian velocity $v_\mathrm{x}$ under the constraint of wind terminal velocity $\bar{v}_\mathrm{w}\sim 100\kms$ for the HH 30 jet, taking the mean terminal velocity $\bar{v}_\mathrm{w} \rightarrow v_\mathrm{x} \sqrt{2\bar{J}_\mathrm{w}-3}$.
We estimate the ranges of $R_\mathrm{x}$ and the mass-loss rates by varying a factor of 2 around a typical dipole moment $\mu_\ast \sim10^{37}$ G\,cm$^{-3}$, with a range of disk accretion rates $\dot{M}_\mathrm{D}$.  A range of $R_\mathrm{x}$ between $0.06$ and $0.19\,\au$ can be obtained with disk accretion rates of $\sim 2$ -- $7 \times10^{-8}\solarmassyr$, yielding $\bar{v}_\mathrm{w}/v_\mathrm{x}$ ranging $1.25$ -- $2.2$ (or, $\bar{J}_\mathrm{w}\sim 2.3$--$4$) within the range of solutions adopted \citep[e.g.,][]{Shu_1994_1,Shang_1998,Shang_2007,Liu_2012}.

Mass conservation yields the mass-loss rate into the wind $\dot{M}_\mathrm{w} = f\dot{M}_{\rm D}$, whereas the rest $(1-f)\dot{M}_{\rm  D}$ goes onto the star at the $\mathrm{X}$-point.
The fraction $f$ is determined by angular momentum balance at the inner edge of the disk, given by the angular momentum removed by the wind from the disk and the excess angular momentum transferred from the star to the disk, which in the $\mathrm{X}$-region gives $f \approx 1/\bar{J}_\mathrm{w}$.

Subsequently a range of mass-loss rates from both hemispheres, $\dot{M}_\mathrm{w} \sim 5\times10^{-9}$ -- $3\times10^{-8} \solarmassyr$ can be obtained.
These naturally cover the inferred values based on the optical emission line intensities, $\sim 1$ -- $4\times10^{-9}\solarmassyr$ \citep[e.g.,][]{Bacciotti_1999,Coffey_2008} for the one-sided jet, which corresponds to the inner denser portion of wide-angle X-wind, occupying about $1/2$ -- $1/4$ of the total mass loss in one hemisphere. These values are within the ranges of disk truncation radii allowable for the combined stellar magnetic dipole $\mu_\ast$, disk accretion rate $\dot{M}_\mathrm{D}$, given the stellar mass $M_\ast$.

The combined physical and kinematic characteristics of the optical emission lines resemble those found in an atomic jet-bearing wide-angle magnetocentrifugal wind as demonstrated in \citet{Shang_1998,Shang_2002,Shang_2010} and \citet{Liu_2012}. 
For example, the predicted ionic neon forbidden lines, known as probes to X-ray ionization, explain the detection of the optical [\ion{Ne}{3}] line in the microjets of DG Tau and Sz 102 \citep{Liu_2016,Liu_2021}. 
The kinematics and optical forbidden emission line intensities of the RW Aur A jets down to $0\farcs1$ ($\sim14\au$) scale in HST/STIS can be reproduced by X-wind models from a launching radius of $0.11\au$ \citep{Liu_2012}.
The diverging streamlines of the wide-angle wind give rise to the extrema of velocities pointing toward and away from the observer. The narrowing of line widths along the jet axis arises as the wind streamlines gradually collimate. These characteristics can be fully accommodated by the kinematic structures and physical conditions of an X-wind jet \citep{Shang_1998,Shang_2002,Shang_2010}, or an extremely inner disk wind resembling an X-wind \citep{Shang_2023}.

\subsection{\texorpdfstring{$^{12}$CO}{12CO} as Molecular Gas Mixed from the Surrounding Ambient Medium}
\label{section:CO as Mix molecular gas}

The occurrence of $^{12}$CO within the shell and envelope-like structures poses a challenging question for its origin. The molecular gas is likely entrained and carried in by the finger-like arcs identified in $^{12}$CO channel maps. These arcs converge toward the jet axis. This $^{12}$CO emission displays clear layered onion-like structures in morphology and kinematics, with the highest line-of-sight velocities from the innermost layers, decreasing outwards on the channel maps as seen in Figure \ref{fig:COChMap_HSTImg}. The inner and smaller shell-like structures appear to be able to close on the axis, while the slower and larger ones tend to leave the axial regions open. The observed velocity of these structures only ranges up to $\sim7\kms$, significantly lower than the EHV jet. However, the velocity differences of a few $\kms$ between the inner and outer molecular layers are clearly shown. This type of pattern has recently been reported in a few Class 0-I sources, some of which were attributed to a slow wind of disk origin launching a considerable distance outward from the inner edge \citep[e.g., HH 212, DG Tau B, and TMC1A;][]{Lee_2021,deValon_2022,Bjerkeli_2016}.

The primary wind is mostly atomic, as demonstrated by the optical jet. Considering the estimated temperature of $\sim 10^{4} \K$, the $^{12}$CO molecular emission is unlikely to originate in the same volume. The $^{12}$CO molecules can not survive in a flow of such physical conditions. The presence of the CO-void region on the transverse PVDs (Figure \ref{fig:TPV_m0_and_clips}) on the axis indicates an absence of molecular $^{12}$CO within the inner outflow cavities. On the other hand, the emission in the highest line-of-sight velocity channel (Figure \ref{fig:COChMap_HSTImg}) moving forward and backward along the line of sight shrinks down to a small region aligned with the axis, suggesting that the highest-speed molecular flow should be in the direction perpendicular to the jet along the line of sight, rather than along the EHV jet direction. This indicates that the source of the momentum, as conveyed in Figure \ref{fig:COChMap_HSTImg}, is the wide-angle portion of the primary wind rather than the jet because there is no major pile-up of matter in the direction of the optical jet.

The mechanism giving rise to the $^{12}$CO emission differs from that observed in the EHV component. If $^{12}$CO had originated from the primary jet within the wind, it should be associated with the EHV component, the $^{12}$CO bullets observed in some young sources. Such molecular emission is missing in the ALMA data of HH 30. Some optical knots observed in the EHV component could have originated in episodic ejections in the primary wind. In contrast, the $^{12}$CO shell-like structures appear to be the edge of material dragged inward from the boundaries of the cavities. They appear to originate from the surrounding ambient region and follow an advection pattern deeply into the outflow lobe. As we will discuss in Section \ref{section:HH30_illustration}, these arc-like filamentary structures resemble those shown in \citet{Shang_2020,Shang_2023}, which argues an origin in mixing ambient material into the wind-blown cavities. Enhanced resolution observations in optical and radio wavelengths will be necessary to elucidate the connection between the $^{12}$CO and optical EHV knots in HH 30. 

In the transverse PV diagrams shown in Figure \ref{fig:TPV_Sketch of ellipses}, the $^{12}$CO emission features two oval-shaped structures crossing each other perpendicularly: the thick green horizontal oval and the vertical oval in red. The thick green oval covers the velocity range, which sums up the arcs of several $\kms$,at $\sim\pm5\kms$ with a width of $\sim2\kms$.
Combinations of shell structures in different velocity gradients can generate these crossing ovals, as demonstrated in \citet{Shang_2023}, using a two-shell schematic model. The blue and red contours, representing the inner higher-velocity and outer lower-velocity parts, are linked by the intermediate green contours in various ways. The vertically extended (in physical position across the outflow lobe) $^{12}$CO structure close to the systemic velocity corresponds mainly to the vertically stretched rings in the PV space shown.

By using the unified wind model, shapes and PV diagrams have been successfully simulated in the wide range of cases presented in \citet{Shang_2020,Shang_2023}, and compared them to existing observational systems, including this work to HH 30. Formulas that trace the outflow shapes in regular space and PV-space have been presented in these works. The momentum-conserving curve has been shown by \citet{Shang_2020,Shang_2023} to approximately trace the interface between compressed wind and compressed ambient field, corresponding to the LV green ring in this work (Figure \ref{fig:TPV_Sketch of ellipses}), the place where the material is entrained from the ambient envelope surrounded by the ambient magnetic field, into the compressed wind region. In the momentum-conserving regime the formula for the outflow shape in PP- and PV-coordinates is straightforward, namely $v_s=v_\mathrm{w}/[1+(\rho_\mathrm{w}/\rho_\mathrm{amb})^{-1/2}]$, where $v_\mathrm{w}$ is the underlying wind velocity, and $v_s$ is the surface expansion velocity. Their ratio is given by the angle-dependent density ratio \citep[Equation 12 of][]{Shang_2023}.
This formula gives the whole range of $v_s$ in the poloidal direction along the surface for the whole lobe.
In our Figure \ref{fig:TPV_Sketch of ellipses}, a line-of-sight (LOS)-averaged velocity of $\sim 5\kms$ is fit to the PV ellipses, using the methods of \citet{Shang_2023} for transverse PV diagrams.
In our framework, the wind velocity is known as $\bar{v}_\mathrm{w}=100\kms$. The measurements of Figure 8 give the observed value of the $\bar{v}_{s,\varpi}\approx5\kms$, which allow estimating the total $\bar{v}_{s}$ as an angle-dependent value between $5\kms$ and $\sim10\kms$, depending on height according to the surface-velocity formula. Applying this formula to these values give $(\bar{v}_s/\bar{v}_\mathrm{w})^2\approx(\bar{\rho}_\mathrm{w}/\bar{\rho}_\mathrm{amb})\approx1/400$--$1/100$, which is appropriate for the system.

The very elongated red oval covers the gas properties only slightly faster than the ambient cloud ($\sim \pm 1 \kms$ to $v_\mathrm{sys}$). It extends beyond $\sim\pm10\arcsec$ across the jet axis, covering an extended length of the outflow lobe and more. This elongated oval is consistent with the cases shown in Figures 18(a), (c), (e) of \citet{Shang_2023} in the presence of the ambient magnetic field. This key feature provides a unique diagnosis of the ambient poloidal field. It suggests that the HH 30 is surrounded by a magnetized environment, as described by the left case of Figure 1 in \citet{Shang_2023}. 

It is important to highlight that our analysis of $^{12}$CO emission and optical observations has revealed the presence of at least three independent ellipse-shaped structures. These structures extend beyond the studied region of \citet{Louvet_2018}, reaching up to $\sim6\arcsec$ above the disk. Multiple droplet-shaped finger-like arcs, connecting from the A0 knot to positions higher than $2\arcsec$ above the disk, have been detected in the $^{12}$CO emission channel maps. These structures suggest multiple layers of shell-like features, which a single conical flow cannot reproduce.

Notice that these features are observationally more salient at lower latitudes, hinting at a wide-angle explanation as more natural than a bowshock explanation. Therefore, the entrainment most probably does not originate from the jet bowshock, while it can well originate from wide-angle features as in our description based on the wide-angle winds of \citet{Shang_2023}.

\subsection{HH 30 as a Clear Case of Magnetic Interplay}
\label{section:HH30_illustration}

\begin{figure*}[ht!]
\plotone{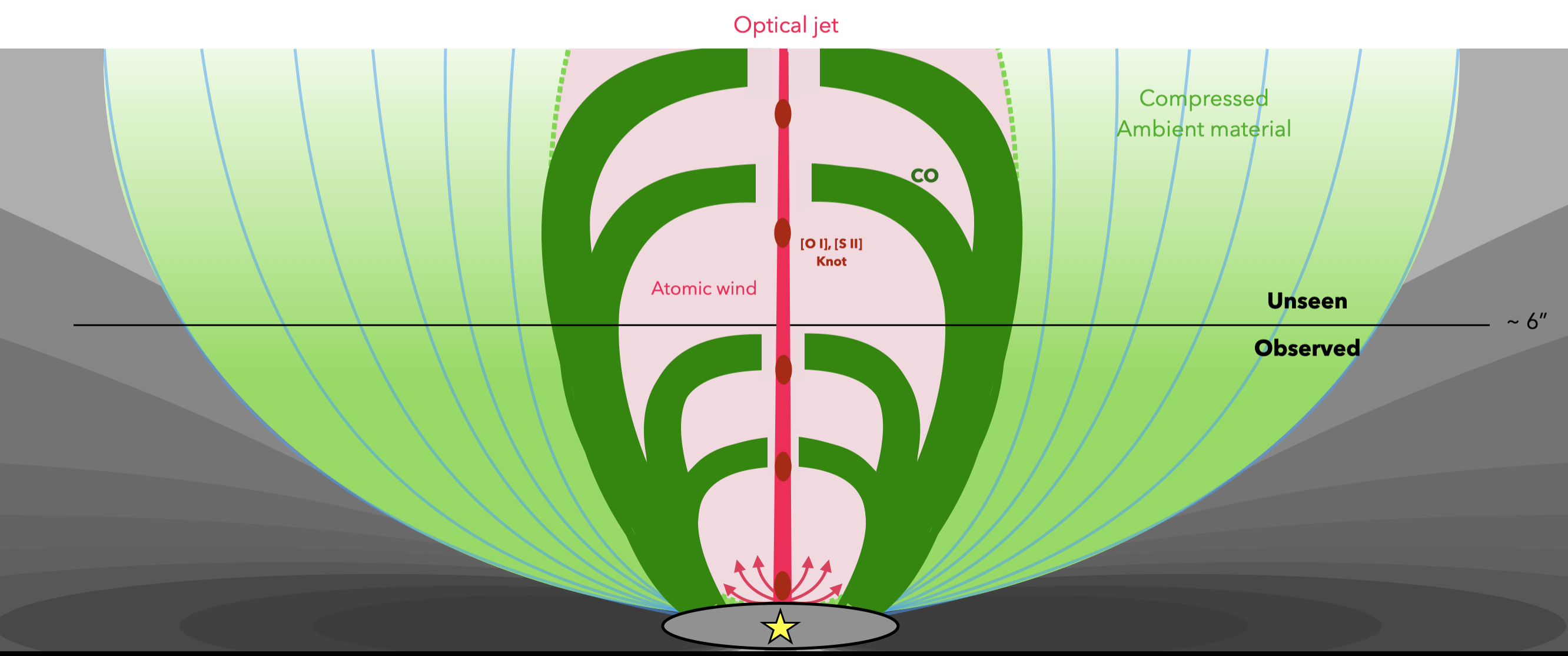}
\caption{
A schematic cartoon illustrating the structures of the HH 30 system. The red area shows each of the components observed in optical, including the wide-angle wind feature at the base of the jet. The distribution of $^{12}$CO material is represented in green, with variations in brightness indicating differences in density. The mixing region and outer part of compressed ambient material are dense enough to be detected. The dashed curve separates the atomic wind and ambient regions. Threaded in the light green regions of the compressed ambient material are the ambient poloidal magnetic field lines in light blue. 
\label{fig:Cartoon}}
\end{figure*}

In this subsection, we discuss the evidence of HH 30 being the example of the magnetic interplay between a wide-angle wind and its ambient surroundings. The schematic picture shown as Figure \ref{fig:Cartoon} illustrates the kinematic structures of the HH 30 outflow system based on the combined observational results from the optical and radio wavelengths. It shows the relative spatial distribution between the atomic/ionic species traced by the optical forbidden emission lines and the molecular species traced by $^{12}$CO line emission.

The reddish area displays each of the components observed in optical, including the wide-angle wind feature at the base of the jet. The distribution of the $^{12}$CO material is represented in dark and light green, with variations in brightness indicating differences in density. The mixing region is mixed with the atomic and partially ionized gas in light pink and the shocked molecular $^{12}$CO in dark green color. The outer edge of the compressed ambient material is the interface between the compressed ambient magnetic field and the uncompressed magnetized envelope region. The dashed curve separates the wind and ambient regions. Threaded in the light green regions are the ambient poloidal magnetic field lines in light blue. The dark green arcs extended from the wind-ambient interface within the wind cavity. The innermost region of the outflow is filled with the atomic wind, shown as the pink-shaded area. Its wide-angle nature, as evidenced by the large line widths in optical lines, is illustrated by the diverging vectors close to the driving source. The axial concentration indicates the observed collimated jet feature as the densest portion in the wind, and a knotty structure is also shown. The ambient material is dragged into the shocked atomic wind region via the finger-like arcs by the action of the KHI and amplified by magnetic forces in the compressed-wind cavity \citep{Shang_2020, Shang_2023}. 
The extended $^{12}$CO arcs manifest the interplay of a wide-angle magnetized wind with its magnetized ambient surroundings. 

The jet driven by an atomic wide-angle wind may be characteristic of the more evolved Class I and II, for which the X-wind framework has been developed. As in HH 30, several other Class I and early Class II YSOs are known to possess atomic jets in optical or infrared forbidden emission lines and are surrounded by nested outer molecular structures. DO Tau is an example source close to a face-on orientation \citep[inferred inclination ranging from $19\arcdeg$ to $28\arcdeg$,][]{Long_2019,Fernandez-Lopez_2020} which shows a pair of infrared [\ion{Fe}{2}] jets with a single high-velocity component \citep[$\sim-100\kms$ on the red and $\sim+170\kms$ on the blue,][]{Erkal_2021} encompassed by a series of ring structures detected in $^{12}$CO by ALMA \citep{Fernandez-Lopez_2020,Huang_2022}. HH 46/47 is another example, showing multiple ring structures \citep{Zhang_2019} surrounding an optical/infrared jet collimated by a wide-angle wind \citep{Nishikawa_2008,Garcia-Lopez_2010}. HL Tau has a pair of collimated optical/infrared jets surrounded by wide-angle molecular hydrogen outflows \citep[][]{Pyo_2006,Takami_2007,Beck_2008}, and recent ALMA CO observations show multiple eccentric ring-like structures similar to DO Tau and HH 46/47 \citep{Bacciotti_2023}. 
DG Tau has been known to show multiple velocity components in the optical and infrared jet/wind emission with increasing spatial widths as flow velocities increase \citep{Bacciotti_2000,Agra-Amboage_2014,Takami_2004}. Its CO emission originates from the outer disk as a few $\kms$ flow \citep{Guedel_2018}, generating an onion-like kinematic structure. RU Lup may be another future candidate for investigating the nested structure of its jet-outflow system. It shows a strong high-velocity peak at $\sim-150\kms$ in the optical [\ion{O}{1}] line indicative of a spatially unresolved microjet from a very inclined orientation \citep{Banzatti_2019}. RU Lup may exhibit a multiple-ring structure composed of outflowing features for $^{12}$CO within a $\sim260$-au scale \citep{Huang_2020}. DG Tau B has been known to have collimated jet features in optical forbidden lines with large line widths at the base of the flow \citep{Podio_2011}, surrounded by conic $^{12}$CO outflows \citep{Zapata_2015,deValon_2020}. A collimated [\ion{Fe}{2}] jet and a wide-angle H$_2$ cavity have recently been detected in TMC1A by JWST in near-infrared \citep{Harsono_2023}, which manifest the properties anticipated for an X-wind jet irradiated and ionized by high-energy photons produced by star--disk activities in the close vicinity to the young star. 

For the younger and more embedded Class 0 sources, the primary jet/wind can be mainly molecular and observed in high-density tracers such as SiO \citep[][]{Glassgold_1991}. The SiO high transitions $J=8-7$ require sufficiently high density $n_\mathrm{H_2}> 10^6\numden$ and a temperature of a few hundred degrees to emit brightly, which places a very tight constraint on the excitation and formation. X-wind characteristics have been identified by the large line widths of SiO as seen in HH 211 \citep{Jhan_2021} and HH 212 \citep{Lee_2022}, and by rotation signatures identified in HH 211 \citep[][SMA]{Lee_2009} and HH 212 \citep[][SMA+ALMA]{Lee_2017} through inferred launching radii of $0.03$--$0.05 \au$. Interestingly, JWST captured collimated mid-infrared [\ion{Fe}{2}] and [\ion{Ne}{2}] jets from the Class 0 source IRAS 15398$-$3359 that possess large line widths of $\pm200\kms$, along with H$_2$ possibly tracing a wide-angle outflow cavity \citep{Yang_2022}. B335, with wide-angle infrared cavity walls delineated by a $^{12}$CO outflow and with very compact SiO emission close to the source \citep{Bjerkeli_2019}, shows via JWST a series of shocked CO and H$_2$ emission arcs and knots \citep{Hodapp_2024} and [\ion{Fe}{2}] knots \citep{Federman_2023} along the outflow axis.
Recent JWST near- to mid-infrared emission line maps reveal an increasing number of YSOs with ionic jets in [\ion{Fe}{2}] and/or [\ion{Ne}{2}] surrounded by wide-angle molecular CO and H$_2$ arcs and cavities or scattered light \citep[e.g.,][]{Federman_2023,Duchene_2023}. The findings are consistent with the jet-bearing wide-angle wind scenario as seen in HH 30 and explained in the X-wind framework and others resembling the X-winds.

\subsection{ Disk-Wind Scenarios for the Slower Molecular Components}
\label{section:Louvet}
\citet{Louvet_2018} investigated the kinematics of HH 30 based on the transverse PVD by modeling the CO outflow as a rotating conic flow and fitting each transverse PV cut of the cone with a pair of ellipses. They presented cuts of transverse PVDs up to $\sim5\arcsec$ above the disk plane, but the fitting and analyses were shown only up to the first $1\farcs8$ ($\sim250\au$). The analysis was restricted to one of the two PV-ellipse sets, discarding the few data points that were deemed to correspond to the other set.
They estimated a CO-component rotation $V_\phi$ from the small PV-inclinations of the ellipses they keep.

\citet{Louvet_2018} assume that the detected rotation $V_\phi$ arises from a disk-wind of poloidal velocity $V_p$, whose launching points $r_0$ they estimate based on theory intended for an axisymmetric disk-wind in a steady state. Equations based on the limits of magnetocentrifugal wind theory are used,
\begin{align}
    r\times V_\phi &=\lambda\sqrt{GM_\ast r_0}\label{eq:rvphi}\\
    V_p &=\sqrt{2\lambda-3}\sqrt{GM_\ast/r_0}\label{eq:vp}
\end{align}
where $r$ is the cylindrical radial coordinate, $M_\ast=0.45\msun$ is the stellar mass \citep{Pety_2006}, and $\lambda$ is the lever-arm ratio parameter, constrained by $\lambda>1.5$ for magnetocentrifugal wind launching \citep[using in this paragraph notations from][]{Louvet_2018}.
The inclinations of their retained PV-ellipses give them an estimate of $r\times V_\phi=38\pm15\au\kms$, and a procedure based on trigonometric guesses about the value of $V_z$, gives them an estimate for $V_p=9\pm1\kms$. They solve the non-linear coupled Equations (\ref{eq:rvphi})--(\ref{eq:vp}) with a graphical method, which gives them a wide range, $0.5$--$2.5\au$, for $r_0$, and a narrow range for $\lambda=1.6$, extremely close to the minimum of $1.5$ for a nonzero $V_p$ mathematical solution, and very sensitive to the trigonometric guessing based on a conical assumption that justifies their value of $V_p$, and the specific value of $\theta=35\arcdeg$ used in obtaining it. These $\lambda$ values lead to a very sub-Keplerian terminal velocity of the hypothetical disk wind.

An alternative analysis of the HH 30 ALMA CO data has been made by \citet{Lopez-Vazquez_2023} by fitting multiple shells to the kinematics of potential disk winds. The multiple shell structure is interpreted as an episodic disk wind ejected intermittently. Three layers are identified, with a constant expansion speed of $4$ to $6\kms$ and a small rotational velocity of $0.5\kms$ at dynamical ages of $500$, $310$, and $260\yr$.
Similar to \citet{Louvet_2018}, they use the \citet{Anderson_2003} formulas to obtain launching points.
An updated launching point value with an upper bound of $4\au$ and a low lever-arm parameter of $1.6$ to $1.9$ is estimated. Still, the inferred terminal MHD disk winds remain sub-Keplerian. The assumption of episodicity, furthermore,
challenges the \citet{Anderson_2003} steady-state conditions in formulating a self-consistent application.

These two works interpret the ``slower'' molecular flows present as originating from an extended disk wind by assuming that an MHD wind launched from larger radii must drive the observed phenomena. The slight asymmetry in the transverse PV diagrams is attributed to the angular momentum carried to the shell locations from the disk. 

Due to the inevitable formation of extended shocked regions inside these outflow systems, analyzing observational data using models requires caution as we discuss next, in Section \ref{section:diff_dw}.

\subsection{Difficulties with Disk-Wind Scenarios for the Slower Molecular Components}
\label{section:diff_dw}

Disk-wind scenarios intended to explain CO outflows \citep[such as][]{Louvet_2018,Lopez-Vazquez_2023} and estimate their launching points, episodic or not, share a major assumption: tracing apparent shell-like molecular structures that exhibit velocities in channels that are significantly lower than the expected jet velocities. One approach \citep[e.g.,][]{Pascucci_2023} interprets these phenomena as the existence of separate slower-moving disk winds launched from somewhat outer regions of the inner disks given by their respective (sub-)Keplerian velocity derived from Equation \ref{eq:vp}, leaving multiple internal shells as material ejected from the disk.

Analysis methods such as \citet{Anderson_2003} attempt to apply either conservation laws or terminal properties to infer disk-wind launching points. The conditions of validity require a pristine and continuous wind in an ideally axisymmetric steady state without episodicity. Exercising these methods requires stringent and observationally inaccessible conditions. Realistic winds interact with their environment, unavoidably driving a magnetized bubble structure that propagates in space \citep{Shang_2023}. This bubble structure's outer, denser, and wider portions correspond to the shocked wind and ambient media modified and deflected by their mutual interplay. Only the innermost portion of the bubble may correspond to the conditions of safe, valid use. Applying conservation laws to deflected and modified flow features will lead to errors in inferring rotation velocities and locating the launching points, with likely overestimates and misleading results.

The formulation by \citet{Anderson_2003} is sensitive to the input rotation velocity inferred from observational data. This is particularly severe in the case of a very low-velocity wind, such as hypothesized by, e.g., \cite{Louvet_2018} and \citet{Lopez-Vazquez_2023}. Their adopted complex methods are subject to wide error ranges. The PV diagrams are fitted to arcs of ellipses corresponding to various shells, which may involve data selection for each shell in a step incurring hard-to-quantify estimation errors. The method has risks of overfitting, especially where ellipse arcs cross or terminate due to sensitivity limits. One important feature of the ellipse model is the small skew in PV space, which is used to obtain the (always small) parameter of rotation $v_\phi\ll v_p$, though with wide ranges of statistical errors as reported by, e.g., \ \citet{Louvet_2018}. Launching point information is then inferred from these values of $v_\phi$. This inference requires caution because of the wide range of statistical and estimation errors and the problematic use of formulas \citep[e.g., those in][]{Anderson_2003} outside their conditions of safe validity.

Most of these disk-wind fittings require extremely low values of the lever-arm parameter $\lambda$, such as $\lambda=1.6$ \citep{Louvet_2018} or $\lambda=1.6$--$1.9$ \citep{Lopez-Vazquez_2023}. These values correspond to a sub-Keplerian terminal wind velocity ($\sqrt{0.2}v_\mathrm{K}$ for $\lambda=1.6$). Such sub-Keplerian velocities, $\lambda<2$, are driven primarily by pressure gradients from weakly ionized and magnetized disks \citep{Pascucci_2023}, strictly not a true cold magnetocentrifugal wind ($\lambda\geq2$) in the sense of \citet{Blandford_1982}. Combined with the wind-driven accretion hypothesis, these small $\lambda$ values may be prone to overestimated mass-loss rates. The presence of other angular-momentum transport mechanisms, MRI, and turbulent viscous accretion give very different mass-loss rates and launching physics.

Our quantitative unified framework is constructed only from physics-based quantitative elements. A unified jet and wind launching theory approach (the X-wind model) is demonstrated in Section \ref{section:Atomic wind}. A theory of outflows, complex interactions between the winds (including the jet) and the ambient medium is demonstrated by \citet{Shang_2006,Shang_2020,Shang_2023} with full numerical results, published as a rich and detailed grid parameterized by wind and ambient medium properties. Useful simplified forms are presented by \citet{Shu_1991} for thin shells and by \citet{Shang_2023} for thicker shells. The relative simplicity of HH 30 allows a very direct quantitative estimate of the velocity range in PVs and qualitative comparisons based on the numerical grid.
Models to map those structures into PV-space are shown by \citet{Shang_2023} and this work.

In this work, the observed molecular features are natural outcomes of the X-wind interaction with an environment. CO emissions in our framework trace the interaction with the external environment, naturally as entrainment features and mixture in the outer parts, demonstrated quantitatively and numerically within a 
broad parameter space constructed in our unified framework.
This framework provides an intuitive explanation for the chemical differentiation observed, making CO features more salient in the outer parts of the outflow. The entrainment in our framework proceeds from magnetized KHI naturally arising in the interplay between wind and ambient surroundings, a universal mechanism connected with bubble structure, quantified and tested numerically by \citet{Shang_2020,Shang_2023}.
Thus, our model explains both observed CO and atomic features in one unified framework based on magnetocentrifugal wind theory.

\section{Summary}
\label{section:summary}

We reexamined the morphological and kinematic properties of the northeastern jet and outflow from the class I protostar HH 30 using the archival HST and ALMA data. The optical HST data reveal the wide-angle-wind nature of the atomic jet through the forbidden emission lines. In contrast, the ALMA data of $^{12}$CO represent the molecular ambient material brought into the outflow lobe by the interplay between the wide-angle wind and its ambient medium. The combined data analysis reveals the nested structures in the HH 30 jet-outflow system.

The optical appearance of HH 30 is characterized by its highly collimated jet feature in the forbidden line imaging, large radial velocity from proper motions ($v\sim100\kms$), and the large line widths ($\Delta v\sim200\kms$) observed in the optical forbidden line spectra. The line diagnostics suggest a jet temperature up to $10^4\K$ and an ionization fraction of $\lesssim10\%$. These combined properties suggest that the HH 30 drives a predominantly atomic and partially ionic wide-angle wind with axial density collimation, which resembles an X-wind jet. The X-wind arises within $0.19$ -- $0.06\au$, for accretion rates of $\sim 2$ -- $7 \times10^{-8}\solarmassyr$, and their respective mass-loss rates from both hemispheres, $\dot{M}_\mathrm{w} \sim 5\times10^{-9}$ -- $3\times10^{-8} \solarmassyr$.    

The molecular appearance of the HH 30 outflow is characterized by the droplet-shaped finger-like arcs observed in $^{12}$CO channel maps and the two distinct ellipses crossing each other on the transverse PVDs. On transverse PVDs, the velocity range of VLV ambient ellipse is $\sim\pm1\kms$ from $v_\mathrm{sys}$ and extends up to $\sim\pm10\arcsec$ whereas the LV ellipse is seen to be confined within $\sim\pm2\arcsec$ with velocity ranging $\sim\pm5\kms$ (with a width of $\sim2\kms$) from $v_\mathrm{sys}$. These LV ellipses are consistent with the predicted feature velocity formula of the outflow surface enclosing the compressed wind by the compressed ambient magnetic field lines in the momentum-conserving regime as shown in \citet{Shang_2020,Shang_2023}. The distinction and crossing of the main ellipses is a signature suggesting that HH 30 possesses an extended ambient magnetic field.

The combined optical and radio analyses help to elucidate the nested morphological and kinematic structure of the HH 30 outflow. The primary atomic wind possesses a collimated jet, representing the innermost optical oval on the transverse PVDs, protruding in velocity from within the CO-void of the LV ellipse. The LV ellipse corresponds to the multiple finger-like converging structures towards the axis. However, given the different properties between the optical and molecular species, it is improbable that $^{12}$CO originates from the inner atomic primary wind. It is more plausible that it is connected to the molecular gas mixed from the surrounding ambient area. This mixing is mainly driven by the KHI, induced by the contrasting characteristics of the two regions: the primary wind and the ambient envelope. Outside of the mixing region, the compressed ambient region, which corresponds in PV to the VLV ellipse, encompasses all elements of the HH 30 outflow and is threaded by a magnetic field. This magnetic influence may restrict the outward expansion of the primary wind. These observed features align well with the predictions of the unified wind framework of \citet{Shang_2020, Shang_2023}.

\vspace*{1cm}
\noindent
The authors acknowledge grant support for the CHARMS group from the Institute of Astronomy and Astrophysics, Academia Sinica (ASIAA), and the National Science and Technology Council (NSTC) in Taiwan through grants 111-2112-M-001-074-, and 112-2112-M-001-030-. The authors acknowledge the access to high-performance facilities (TIARA cluster and storage) in ASIAA\@. 
D.J.\ is supported by NRC Canada and by an NSERC Discovery Grant.

The HST/STIS spectra used in this work were obtained from the Mikulski Archive for Space Telescopes (MAST) at the Space Telescope Science Institute (STScI) under General Observing Program 9164, and can be accessed via \dataset[DOI: 10.17909/fejm-x413]{https://doi.org/10.17909/fejm-x413}. The HST/WFPC2 images can be accessed via \dataset[DOI: 10.17909/jxke-sz23]{https://doi.org/10.17909/jxke-sz23}, and were obtained from the Hubble Legacy Archive (HLA), which is a collaboration between the Space Telescope Science Institute (STScI/NASA), the Space Telescope European Coordinating Facility (ST-ECF/ESA) and the Canadian Astronomy Data Centre (CADC/NRC/CSA)\@. STScI is operated by the Association of Universities for Research in Astronomy, Inc., under NASA contract NAS 5-26555. 
This work makes use of the following ALMA data: ADS/JAO.ALMA\#2013.1.01175.S\@. ALMA is a partnership of ESO (representing its member states), NSF (USA), and NINS (Japan), together with NRC (Canada), NSTC, and ASIAA (Taiwan), as well as KASI (Republic of Korea), in cooperation with the Republic of Chile. The Joint ALMA Observatory is operated by ESO, AUI/NRAO, and NAOJ\@. 
This research has made use of SAO/NASA Astrophysics Data System.

\vspace{5mm}
\facilities{ALMA, HST (STIS, WFPC2)}

\software{Python3 \citep{10.5555/1593511}, 
          Numpy \citep{Numpy},
          Astropy \citep{astropy2013,astropy2018,astropy2022},  
          Scipy \citep{2020SciPy-NMeth}, 
          OpenCV \citep{opencv_library},  
          SpectralCube \citep{Spectral_cube},
          Matplotlib \citep{hunter2007},
          pySpecKit \citep{pyspeckit}
          }

\bibliography{HH30}{}
\bibliographystyle{aasjournal}



\end{document}